\documentclass{nws}
\usepackage{graphicx}
\usepackage{natbib}
\usepackage{graphicx}
\usepackage{amsfonts}
\usepackage{subfig}
\usepackage{makecell}
\usepackage{tikz}
\usepackage{url} 

\title[Faster MCMC for Gaussian Latent Position Network Models]
      {Faster MCMC for Gaussian Latent Position Network Models}

 \author[N.A. Spencer, B.W. Junker, and T.M. Sweet]{Neil A. Spencer\\ 
 Harvard University\\
 nspencer@hpsh.harvard.edu\\
 \and \; Brian W. Junker\\ Carnegie Mellon University\\ 
 brian@stat.cmu.edu \\
 \and \; Tracy M. Sweet\\ University of Maryland College Park\\ 
 tsweet@umd.edu
 }

\begin{document}

\label{firstpage}

\maketitle
\begin{abstract}
Latent position network models are a versatile tool in network science; applications include clustering entities, controlling for causal confounders, and defining priors over unobserved graphs. Estimating each node's latent position is typically framed as a Bayesian inference problem, with Metropolis within Gibbs being the most popular tool for approximating the posterior distribution. However, it is well-known that Metropolis within Gibbs is inefficient for large networks; the acceptance ratios are expensive to compute, and the resultant posterior draws are highly correlated. In this article, we propose an alternative Markov chain Monte Carlo strategy---defined using a combination of split Hamiltonian Monte Carlo and Firefly Monte Carlo---that leverages the posterior distribution's functional form for more efficient posterior computation. We demonstrate that these strategies outperform Metropolis within Gibbs and other algorithms on synthetic networks, as well as on real information-sharing networks of teachers and staff in a school district. 
\end{abstract}

\tableofcontents

\newpage

\section{Introduction}
\label{sec:intro}

Network data---measurements of relationships across sets of entities---are becoming increasingly common across science and industry, largely due to technological advances in data collection and storage. Common sources of network data include social networks \citep{carrington2005models}, citation networks \citep{ji2016coauthorship}, gene regulatory networks \citep{hecker2009gene}, disease transmission networks \citep{newman2002spread}, neural connectomes \citep{chen2016joint}, transportation networks \citep{xie2009modeling}, and food webs \citep{chiu2011unifying}. 
A broad range of statistical tools based on stochastic graphs \citep{goldenberg2010survey, crane2018probabilistic} are available for probabilistically modeling networks, ranging from the simple Erd\H{o}s-Renyi model \citep{erdHos1960evolution} to sophisticated latent variable models \citep{airoldi2008mixed, clauset2008hierarchical, fosdick2018multiresolution, dabbs2020conditionally}. Latent variable models can be defined to capture common network properties such as community structure, hierarchical structure, and degree heterogeneity. 

Evaluating the likelihood of a latent variable model has a computational complexity that is quadratic in the number of nodes. These models are thus costly to fit to large networks, especially if one wishes to quantify uncertainty in a Bayesian modeling and inference framework \citep{gelman2013bayesian}. For instance, traditional Markov chain Monte Carlo algorithms \citep{gamerman2006markov} such as Gibbs sampling or random walk Metropolis can require tens of thousands of likelihood evaluations to accurately quantify expectations and uncertainties. This computational burden is even larger when the chains are slow-mixing, which is often the case for Bayesian hierarchical models.

In this work, we develop a faster Markov chain Monte Carlo algorithm for a class of latent variable network models called the latent position network model (LPM). LPMs---originally proposed by \cite{hoff2002latent}---have been applied to a variety of statistical problems, including modeling network interventions \citep{sweet2013hierarchical}, clustering entities \citep{handcock2007model}, modeling social influence \citep{sweet2020latent}, controlling for causal confounders \citep{mcfowland2021estimating}, and defining priors on unobserved graphs \citep{linderman2016bayesian}. Each node in a LPM possesses a real-valued latent variable (its \emph{position}), with each edge treated as an independent Bernoulli random draw depending on the participating nodes' latent positions. These probabilities are modeled as a decreasing function of the nodes' latent distance, thus promoting homophily and triadic closure (i.e. a friend of a friend is more likely to be a friend). Edge probabilities may also depend on covariates, such as whether the entities share a common observed trait.

The principal task in fitting a LPM is to infer the latent positions, as well as the parameters of the link function. In a Bayesian modeling and inference framework, the posterior distribution of these parameters quantifies uncertainty in the corresponding estimates. Evaluating and summarizing this posterior distribution requires intensive computation, namely because of an intractable normalization constant.

The standard tool for computing posterior summaries has been Markov chain Monte Carlo (MCMC) via Metropolis within Gibbs \citep{handcock2007model, raftery2012fast}. This technique side-steps explicit computation of the normalization constants, and can approximate posterior expectations arbitrarily well if run long enough. However, accurate inference via Metropolis within Gibbs can be computationally infeasible for large networks, largely due to two problems: (1) The random walk step size required to obtain high acceptance rates shrinks as the number of nodes grows, resulting in slowly mixing chains for large networks, and (2) the computational complexity of performing a full sweep of position updates is quadratic in the number of nodes, so each iteration for a large network is expensive to compute. We address these challenges in this article through the development of a more efficient MCMC algorithm.

We are not the first to recognize these limitations of Metropolis within Gibbs for LPMs. In recent years, multiple approaches for approximating the likelihood have been proposed to scale up Bayesian inference of LPMs to large networks. \cite{raftery2012fast} proposed a case-control based approach, sub-sampling the non-edge dyads to approximate each acceptance ratio in Metropolis within Gibbs. \cite{rastelli2018computationally} proposed a discrete-grid approximation of the latent positions, simplifying each likelihood evaluation. \cite{salter2013variational} proposed variational inference as an alternative to MCMC. Though each of these approaches speeds up posterior inference, the improvements come at the cost of biasing the results with the likelihood approximations. 

Our approach is instead based on Hamiltonian Monte Carlo (HMC). HMC \citep{duane1987hybrid, neal2011mcmc, betancourt2017conceptual} and its variants \citep{girolami2011riemann, hoffman2014no, betancourt2016identifying} are a class of MCMC algorithms that leverage Hamiltonian dynamics to construct gradient-informed proposals for differentiable posterior distributions. Well-tuned HMC proposals produce large moves while maintaining high Metropolis-Hastings acceptance rates. HMC can thus be much more efficient than traditional random walk-based methods, especially in high dimensions, without introducing any bias in the likelihood.

In recent years, the use of HMC algorithms has been democratized in the software Stan \citep{carpenter2017stan}. Stan implements a specialized version of HMC that generally applies across a broad class of Bayesian models, with built-in tools for diagnosing Markov chain mixing problems. However, Stan's generality requires limiting its flexibility, such as requiring all discrete latent variables to be marginalized, and the rest to be updated simultaneously. Usually these are small sacrifices for easy HMC implementation with built-in mixing diagnostics. However, MCMC for large LPMs often stretches one's computational resources to their limit. We thus need all tools at our disposal to optimize our inference strategy, including sampling discrete random variables and block updates of variables.

The specialized HMC-based sampling strategy we develop in this article is specifically intended for Gaussian LPMs \citep{rastelli2016properties}, a class of LPMs for which the link probability function decays like a half-Gaussian probability density function. This class of LPMs was originally studied because they are easy to work with analytically. We show here that the Gaussian-inspired link function also provides computational advantages---the log posterior can be split into Gaussian and non-Gaussian components, thus facilitating efficient integration of HMC via split HMC \citep{shahbaba2014split}. Moreover, we further increase the efficiency for sparse networks by developing an exact dyad subsampling scheme based on Firefly Monte Carlo (FlyMC; \cite{maclaurin2015firefly}). This scheme allows us to subsample the non-edge dyads, decreasing the complexity of the non-Gaussian component of the posterior while maintaining an exact MCMC strategy. To complete the LPM fitting algorithm, we also include Markov chain updates for the parameters of the link function. We also extend our sampling strategy to accommodates categorical covariates in the link function, as well as prior dependence between latent positions in the network as in longitudinal latent position models \citep{kim2018review}.

The remainder of the article is organized as follows. Section~2 establishes notation and provides the necessary background information pertaining to LPMs, Gaussian LPMs and Hamiltonian Monte Carlo. Section~3 outlines the ingredients of our new computation methodology for Gaussian LPMs: split Hamiltonian Monte Carlo and firefly Monte Carlo, then combines them with updates to the link function parameters to define a new Markov chain Monte Carlo strategy. Section~4 presents two empirical studies to demonstrate the superiority of our algorithm. Study 1 uses synthetically-generated examples to demonstrate the superior performance of our method compared to a variety of existing approaches in the literature such as Metropolis within Gibbs, elliptical slice sampling, Stan, and the No-U-turn sampler. Study 2 demonstrates the extent to which our algorithms outperform Metropolis within Gibbs for fitting information-sharing models amongst teachers and staff in a school district. Section~5 contains some concluding remarks.

\section{Preliminaries}

The following notation will be used throughout the paper. We use $\mathbb{R}$ to denote the set of real numbers, $\mathbb{R}_+$ to denote the set of non-negative real numbers, $\mathbb{N}$ to denote the set of natural numbers, and $[n]$ to denote the set $\{1, \ldots, n \}$ of natural numbers less than or equal to $n$. For a set $S$, we use $S^{d}$ to denote the collection of all $d$-length vectors with entries from $S$ and $S^{n \times d}$ to denote collection of possible $n \times d$ matrices with entries from $S$. For two sets $S_1, S_2$, $S_1 \times S_2$ denotes their Cartesian product.

For a vector $z \in \mathbb{R}^d$, $z_i$ denotes its $i$th entry and $\|z\|$ denote its Euclidean norm. For a matrix $B \in \mathbb{R}^{n \times d}$, $B_{i \cdot}$ denotes its $i$th row, $B_{\cdot i}$ denotes its $i$th column, $B_{ij}$ denote its $(i,j)$th entry, $B^T \in \mathbb{R}^{d \times n}$ denotes its transpose, and $B^{-1}$ denotes its inverse. We use $I_n$ to denote the $n \times n$ identity matrix. 

We represent networks among $n$ entities as undirected binary graphs on $n$ nodes. We use $A \in \left\{0,1 \right\}^{n \times n}$ to denote the adjacency matrix of the graph, with $A_{ij} = 1$ indicating the presence of an edge between nodes $i$ and $j$, and $A_{ij} = 0$ indicating its absence. Our focus is on undirected graphs, so $A_{ij} = A_{ji}$ for all dyads $(i,j) \in [n]^2$. For simplicity, we use $A$ to refer to both a graph and its adjacency matrix interchangeably, using $[n]$ index the nodes according to the order of their rows in the adjacency matrix. We use the shorthand $E_A \subseteq [n]^2$ to denote the set of edges associated with $A$, and $\left\{(i,j) \notin E_A\right\}$ to denote the set of edges absent from $E_A$. The combinatorial Laplacian of $A$ is denoted as $L^A \in \mathbb{R}^{n \times n}$. Specifically, $L^A = D^A - A$ where $D^A$ is a diagonal matrix of the node degrees $D^{A}_{ii} = \sum_{j=1}^n A_{ij}$.

\subsection{Latent Position Network Models}\label{lpmdefinition}

In the distance-based latent position network model (LPM) of \cite{hoff2002latent}, each node $i \in [n]$ is modeled as having a $d$-dimensional latent position $z_{i} \in \mathbb{R}^d$ for some positive integer $d$ (typically $d=2$ or$d=3$ to facilitate visualization). It is convenient to arrange these latent positions in a matrix $Z \in \mathbb{R}^{n \times d}$, where $Z_{i \cdot } = z_{i}$. The edges $A_{ij}$ are modeled as being generated according to 
\[
\mathbb{P}(A_{ij} = 1|Z) = K(\|z_{i} -z_{j} \|, x_{ij})
\]
where $\|z_{i} - z_{j }\|$ denotes the distance between nodes $i$ and $j$, $x_{ij}$ represents any relevant edge-specific covariates for nodes $i$ and $j$, and $K$ is the \emph{link function}---a non-increasing function from $\mathbb{R}_+ \times \mathcal{X}$ to $[0,1]$. Here, $\mathcal{X}$ denotes the range of possible values of the covariate $x_{ij}$ for each dyad $(i,j) \in [n]^2$. In this article, we assume each covariate $x_{ij}$ is categorical taking on $C$ distinct values. For notational convenience if there are no covariates, we will take $C = 1$ and let $x$ be an $n \times n$ matrix of ones.


In their original version of the LPM, \cite{hoff2002latent} proposed modeling $K$ as a logistic function of the latent distance and the covariate according to
\[
K(\|z_{i} -z_j\|, x_{ij}) = (1 + \exp{(\alpha + \beta_{x_{ij}} + \|z_i - z_j \|)})^{-1},
\]
where the parameters $\alpha \in \mathbb{R}$ and $\beta \in \mathbb{R}^C$ control the total number of edges and the effect of the covariates, respectively. Recently, \cite{rastelli2016properties} proposed an alternative form for $K$ inspired by the functional form of the Gaussian probability density function---aptly named the Gaussian Latent Position Model (GLPM). Their original exposition did not consider covariates, taking the form
\[
K(\|z_i -z_j\|) =  \tau \exp{\left(-\frac{1}{2\gamma^2} \|z_i - z_j\|^2 \right)},
\]
where $\tau \in [0,1]$ controls the number of edges (i.e. sparsity level) and $\gamma^2 >0$ controls the decay of the link probabilities. 

Thus far, two advantages of GLPMs over logistic LPMs have been identified in the literature. The Gaussian-like choice of $K$ yields closed-form expressions for various network statistics of GLPMs which makes them easier to theoretically analyze than logistic LPMs \citep{rastelli2016properties}. The lighter tails of the Gaussian link function are also conducive to proving consistency of the maximum likelihood estimator of the latent positions \citep{spencer2017projective}. In this paper, we identify and explore yet another advantage of GLPMs---the Gaussian shape of $K$ facilitates faster posterior inference techniques. 

Our work considers an extension of the GLPM to accommodate categorical covariates. Specifically, we consider
\begin{equation}
\label{gaussianlink}
K(\|z_i -z_j\|, x_{ij}) =  \tau_{x_{ij}} \exp{\left(-\frac{1}{2\gamma^2} \|z_i - z_j\|^2 \right)},
\end{equation}
with parameters $\tau \in [0,1]^C$ and $\gamma^2 > 0$. Here, the effect of the covariate is encoded in the vector $\tau$, allowing for subnetworks corresponding to certain covariate categories to be sparser than others. This single covariate formulation can be extended without loss of generality to multiple discrete covariates, with or without interactions, by a suitable mapping of the joint range space of covariates into $[C]$. For notational conciseness, we occasionally omit the dependence of $K$ on $x_{ij}$ in this article. Together, the Gaussian shape and factorizability of $K$ can be exploited to speed up Bayesian inference.

Before proceeding, it is important to acknowledge that the parameters $\gamma$ and $Z$ in a GLPM are together identified only up to a multiplicative constant. That is, for any viable estimates $\gamma = \hat{\gamma}$ and $Z = \hat{Z}$, a model defined by 
\begin{equation} \label{breakout}
\gamma = 1, \; \; \; Z = \hat{\gamma}^{-1} \hat{Z}
\end{equation}
has an equivalent likelihood. This second parameterization is known as a centered parameterization \citep{papaspiliopoulos2007general}, which can be computationally advantageous for inferring $\gamma^2$. We exploit it in Section~\ref{gamma2}.

\subsection{Existing Computational Strategies for Bayesian Inference of LPMs}\label{LPMBayes}

Fitting a LPM to a graph $A$ can be separated into two interdependent tasks: (1) inferring the latent positions $Z$, and (2) inferring the parameters of the link function. Depending on the modeling objective of the problem at hand, either (1) or (2) could be the primary inferential target. For instance, $Z$ is the primary inferential target when controlling for causal confounders \citep{mcfowland2021estimating}, but $\tau$ (as in \ref{gaussianlink}) is the primary target when estimating the effect of a covariate on edge probabilities. Regardless, both inference tasks are typically carried out in a single Monte Carlo algorithm; independent priors are placed on both the parameters and the latent positions, and their posterior distribution is approximated with samples draw according to Markov chain Monte Carlo.

To simplify exposition, we will present our strategies for the two inference tasks separately. Here, we review existing computational strategies from the literature for inferring the latent positions $Z$ conditional on $K$. All discussion of inference on the parameters of $K$ for the GLPM is deferred until Section~\ref{learningK}. 

Bayesian inference of the latent positions $Z$ depends on the link function $K$, the observed covariates $x$, and two additional inputs: an observed network (encoded by an adjacency matrix $A \in \left\{0,1\right\}^{n \times n}$), and a prior on $Z \in \mathbb{R}^{n \times d}$. The standard prior choice for $Z$ in the literature has been an independent isotropic $d$-dimensional Gaussian on each row (i.e. latent position) of $Z$. Here, we generalize this prior to $Z_{\cdot k} \sim N(0, \Omega^{-1})$---defined independently with a shared $\Omega$ for each $k \in [d]$. That is, the nodes' positions are independent and identically distributed across dimensions, but can be dependent across nodes within each dimension. Without loss of generality, any Gaussian prior exhibiting dependence of a nodes' position across dimensions can be transformed into an equivalent prior with independence across dimensions via a rotation\footnote{In this sense, the dimensions of $Z$ behave like principal components in principal component analysis}.

This more general set-up for the prior allows for known structural information---such as feature-informed node clustering or temporal dependence---to be included as non-zero entries in the precision matrix $\Omega \in \mathbb{R}^{n \times n}$. Other priors, such as a mixture of Multivariate Gaussians \citep{handcock2007model, krivitsky2009representing}, are beyond the scope of this paper, but would involve a straightforward extension of the methods presented here.

Given the prior $Z_{\cdot k} \sim N(0, \Omega^{-1})$ for $k \in [d]$, the posterior distribution on $Z$ is given by
\begin{equation}
\label{kpost}
\mathbb{P}(Z|A) \propto \prod_{(i,j) \in E_A} K(\|z_i -z_j\|) \prod_{(i,j) \notin E_{A}} \left(1- K(\|z_i -z_j\|)\right) \exp{\left(- \frac{1}{2} \sum_{k=1}^d Z^T_{\cdot k} \Omega Z_{\cdot k} \right)}.
\end{equation}
The normalization constant for this density is a $(n\times d)$-dimensional integral that cannot be computed analytically. Instead, we must rely on approximate methods for calculating expectations with respect to the posterior.

In their seminal LPM paper, \cite{hoff2002latent} proposed for the posterior computation of $\mathbb{P}(Z|A)$ to be carried out via Markov chain Monte Carlo (MCMC). They obtained a Markov Chain $Z^1, Z^2, \ldots, Z^T$ with stationary distribution $\mathbb{P}(Z|A)$ by repeatedly applying a random walk Metropolis update to all latent positions $Z$ simultaneously. As is the case for most MCMC algorithms, ensuring an adequate Metropolis-Hastings acceptance rate requires that the standard deviations of these random walk updates be appropriately tuned using a series of short pilot runs. However, these joint random walk proposals are known to be inefficient when the posterior is high-dimensional (e.g. for networks with many nodes) because the random walk standard deviation required to obtain reasonable acceptance rates is simply too small to explore the space efficiently.

In an effort to alleviate this slow mixing, the subsequent LPM literature (e.g. \citep{handcock2007model, raftery2012fast}) use a Metropolis within Gibbs strategy for updating the latent positions instead. In \emph{Metropolis within Gibbs}, the latent positions $(z_i)_{i \in [n]}$ are updated one at a time in sequence according to a random walk via a symmetric kernel $q$ centered at its current position (e.g. a scaled isotropic Gaussian or multivariate uniform). This approach is implemented in the popular \textsf{R} package \texttt{latentnet} \citep{krivitsky2008fitting}; it still widely used today \citep{fosdick2018multiresolution, aliverti2019spatial, sweet2020latent}.

A sweep of the Metropolis within Gibbs algorithm can be summarized as follows. For each $i \in [n]$, 
\begin{enumerate}
\item Propose $z'_i \sim q_{\delta}(z_i, z'_i)$.
\item Accept this proposal with probability equal to
\begin{equation}
\frac{\mathbb{P}(Z'|A)}{\mathbb{P}(Z|A)} = \frac{\exp{\left(\sum_{k=1}^d Z_{\cdot k}^T \Omega Z_{\cdot k}) \right)}}{\exp{\left(\sum_{k=1}^d (Z'_{\cdot k})^T \Omega Z'_{\cdot k}) \right)}} \prod_{j:(i,j) \in E_A} \frac{K(\|z'_i -z_j\|)}{K(\|z_i -z_j\|)} \prod_{j:(i,j) \notin E_A} \frac{1- K(\|z'_i -z_j\|)}{1- K(\|z_i -z_j\|)} 
 \label{MH1}
\end{equation}
Otherwise reject and keep $z_i$.
\end{enumerate}
The matrix $Z'$ in (\ref{MH1}) is constructed such that $Z'_i = z_i'$ and $Z'_j = z_j$ for all other $i \neq j$. The notation $z'_i \sim q_{\delta}(z_i, z'_i)$ denotes drawing $z'_i$ from a symmetric distribution centered at $z_i$ with $\delta > 0$ denoting a tuning parameter for the width, or \emph{step size} of the proposal. Computing (\ref{MH1}) involves only the prior for $z_i$ and the (at most $n$) likelihood terms corresponding to dyads containing $i$---all other terms are equivalent for $Z$ and $Z'$. For a fully observed network $A$, each full sweep updating $Z$ thus requires $O(n^2)$ computations. 

As with random walk Metropolis, it is standard practice to tune $\delta$ using preliminary tuning runs to achieve a desired acceptance rate\footnote{Empirically, we have found that for LPMs, a Metropolis within Gibbs acceptance rate somewhere between 20 and 30 percent gives optimal results---this is consistent with related optimal scaling theory \citep{roberts2001optimal}}. The required value of $\delta$ typically shrinks as $n$ grows, meaning that chains must be run much longer to achieve mixing when fitting larger networks. For example, Figure~\ref{tuningeps} in Section~\ref{adfigs} of the Appendix demonstrates the decreasing relationship between the number of nodes and the tuned Metropolis within Gibbs step size $\delta$ for the variety of different synthetic networks considered in Section~\ref{exp1}. 

For large enough $n$, approximating the posterior using Metropolis within Gibbs thus also becomes intractable \citep{raftery2012fast}---the step-size is too small to efficiently explore the space given the complexity of computing the Metropolis-Hastings acceptance ratios. There have been multiple recent proposals that approximate the LPM likelihood \citep{raftery2012fast, rastelli2018computationally} to alleviate the computational burden of the accept-reject step. But as noted in the introduction, these approximations introduce non-vanishing bias in the subsequent inference.  

Our goal in this article is to avoid such bias completely by developing an MCMC algorithm that outperforms Metropolis within Gibbs without sacrificing exactness. To accomplish this, we use a Monte Carlo algorithm known as Hamiltonian Monte Carlo (HMC). 

\subsection{Hamiltonian Monte Carlo} \label{HMC}

Hamiltonian Monte Carlo (HMC) is an auxiliary variable MCMC algorithm that uses the gradient of the log posterior to inform an efficient Markov proposal kernel. Inspired by Hamiltonian dynamics, HMC augments the posterior distribution with a ``momentum'' variable for each target parameter, framing the task of proposing the next state as that of simulating Hamiltonian motion of an object along a high-dimensional surface. 

An HMC chain consists of a sequence of snapshots of an object sliding along the frictionless surface. The object's momentum is randomly refreshed after each snapshot, thus resulting in a sequence of stochastic draws. By using the negative log posterior as the energy function to inform its motion, HMC provides large step sizes that nevertheless maintain high Metropolis-Hastings acceptance rates. Moreover, these ratios have closed forms due to the properties of Hamiltonian dynamics (namely reversibility and volume preservation). The algorithm is thus efficient for exploring high-dimensional posteriors. 

General implementations of HMC have recently gained traction in the literature for fitting LPMs within large hierarchical models \citep{linderman2016bayesian, salter2017latent}. In this work, we will develop an HMC algorithm that is specifically tooled for inference in LPMs. We now provide a description of HMC, placing emphasis on the components relevant to the development of our algorithm for the LPM. For more detailed reviews of the theory and practice of MCMC, see \cite{neal2011mcmc} or \cite{betancourt2017conceptual}. 

Consider a target density $p(Z)$ that is differentiable with respect to its real-valued arguments $Z \in \mathbb{R}^d$. HMC targets an augmented version of this density $p(Z, U) = p(Z) q(U)$ where $U \in \mathbb{R}^d$ is a vector of auxiliary momentum variables---each corresponding to an entry in $Z$. Note that because the density $p(Z, U)$ admits $p(Z)$ as a marginal, discarding the $U$'s from a Markov chain targeting $p(Z, U)$ yields draws from $p(Z)$.

Let $q(U)$ be a zero mean multivariate Gaussian density with covariance matrix $M \in \mathbb{R}^{d \times d}$, and let $H(Z, U) =  - \log(p(Z)) - \log(q(U))$. This function $H(Z,U)$ plays the role of energy in the Hamiltonian dynamics of HMC, with the covariance $M$---referred to as the \emph{Mass matrix} \citep{neal2011mcmc} or \emph{Euclidean metric} \citep{betancourt2017conceptual}---controlling the effect of the momentum on the dynamics.

Hamiltonian motion over $H(Z,U)$ is governed by the following differential equations:
\begin{align}
\frac{\textrm{d} Z_i}{\textrm{d} t} &= \frac{\partial H(Z, U)}{\partial U_i} = (M^{-1} U)_i \label{DE1} \\
\frac{\textrm{d} U_i}{\textrm{d} t} &= \frac{- \partial H(Z, U)}{\partial Z_i} = \frac{\partial \log(p(Z))}{\partial Z_i} \label{DE2}.
\end{align}
Here, $(M^{-1} U)_i$ denotes the $i$th coordinate in the vector $M^{-1} U$, and $t$ represents the artificial ``time'' for which the Hamiltonian trajectory is computed. That is, the derivatives of $Z$ and $U$ with respect to $t$ reflect the rate of change in these quantities along Hamiltonian trajectory. Given an initial state $(Z^0, U^0)$, HMC generates a Markov chain of snapshots $(Z^{j}, U^{j})_{j \in \mathbb{N}}$ with stationary distribution $p(Z,U)$ by iterating between simulating Hamiltonian motion for a fixed integration time $T > 0$, then refreshing the momentum according to its conditional distribution. The integration time $T$ is typically specified by the user to control the length of time between momentum updates.

Given $(Z^{j}, U^{j})$, the next draw $(Z^{j+1}, U^{j+1})$ is obtained using the following steps
\begin{enumerate}
\item Update the momentum variables via Gibbs $U^{j'} \sim \textrm{MVN}(0, M)$.
\item Simulate Hamiltonian motion $(Z^{j}, U^{j'}) \rightarrow (Z^{j''}, U^{j''})$ for $T$ time units.
\item Accept the move $(Z^{j+1}, U^{j+1}) = (Z^{j''}, -U^{j''})$ with probability
\begin{equation} \label{MH2}
\textrm{min}\left(\frac{p(Z^{j''}, U^{j''})}{p(Z^{j'}, U^{j'})}, 1\right).
\end{equation}
Otherwise reject the move, letting $(Z^{j+1}, U^{j+1}) = (Z^{j}, U^{j'})$. 
\end{enumerate}
The negation of $U^{j''}$ in Step 3 ensures the proposal is reversible\footnote{In practice, the marginal distribution of $Z$ (and not the joint distribution of $Z, U$) is the target of HMC, so this negation step can be omitted because it is immediately changed by subsequent Gibbs update of $U$ \citep{neal2011mcmc}.}.

The performance of HMC as described above depends on two user-specified parameters: the mass matrix $M$ and the integration time $T$. Before we discuss choosing these parameters, we must first address how to simulate Hamiltonian motion.

If the Hamiltonian motion in Step 2 above were to be simulated exactly, the Metropolis-Hastings correction in Step 3 would be unnecessary because the ratio would be exactly one \citep{neal2011mcmc}. This property is guaranteed by the conservation of energy in Hamiltonian motion---Step 2 simply moves along a density contour of the augmented distribution. Unfortunately, exact simulation of the Hamiltonian motion is not possible for most posterior densities that arise in Bayesian inference---there is no known analytic way to move along the contours. 

In practice, simulation of the trajectory of Hamiltonian motion is typically carried out using approximate numerical integrators of the differential equations, the most popular of which is the \emph{leapfrog integrator} \citep{neal2011mcmc}. The leapfrog integrator discretizes the Hamiltonian motion via alternating linear updates of $U$ and $Z$ until the trajectory of length $T$ has been simulated. The following steps are iterated $L \in \mathbb{N}$ times:
\begin{align*}
U &\leftarrow U - \frac{\epsilon}{2}\frac{\partial H}{\partial Z}(Z,U)\\
Z &\leftarrow Z + \epsilon M^{-1} U\\
U &\leftarrow U - \frac{\epsilon}{2}\frac{\partial H}{\partial Z}(Z,U)
\end{align*}

Together, the user-specified parameters $\epsilon > 0$ (step-size) and $ L \in \mathbb{N}$ (the number of steps) define the integration time $T = L\epsilon$. Smaller values of $\epsilon$ provide more accurate approximations of the Hamiltonian motion, and thus a higher the Metropolis-Hastings acceptance rate. However, they also require correspondingly larger values of $L$---and thus more computation---to simulate a given integration time $T$. It is thus important to strike a balance between the two to obtain adequately high acceptance rates for reasonably correlated draws without wasting computational resources.  

Choosing the user-specified parameters $M$ and $T$ for HMC via leapfrog amounts to choosing three parameters: the mass matrix $M$, the step size $\epsilon$ and the number of steps $L$. The integration time $T = L \epsilon$ is a product of the choices of $\epsilon$ and $L$. 

Like tuning the step-size for traditional Metropolis algorithms, standard practice for choosing $\epsilon$ and $L$ is to conduct preliminary tuning runs at various parameter levels, looking for values that maximize the chain's efficiency. The matrix $M$ can also be chosen this way. However, a more theoretically motivated heuristic \citep{betancourt2017conceptual} is to set $M$ to the precision matrix of the posterior as estimated from the preliminary chains. In practice, the true precision matrix may be dense (and thus expensive to compute), making a diagonal or low-rank approximations \citep{carpenter2017stan,bales2019selecting} more suitable. 

Unfortunately, efficiently tuning all three of $L$, $\epsilon$, and $M$ can itself be a computationally burdensome because the three parameters are interdependent (the optimal choice of $\epsilon$ relies particularly heavily on the choice of $M$). Indeed, under the standard settings, it is typical for Stan to devote more time to adapting $L$, $\epsilon$ and $M$ than running the final Markov chain. When the computational problem is already straining the computational budget at hand---such as in the case of large LPMs---these tuning costs can be prohibitive. We thus seek a simpler, more easily tunable algorithm that is specialized to large LPMs.

Before proceeding, it is worth noting that strategies exist for which $L$, $\epsilon$ and $M$ are not necessarily held constant through all regions of the posterior. For instance, Riemannian HMC \citep{girolami2011riemann} adapts $M$ based on the current state of the chain, and the NUTS algorithm \citep{hoffman2014no} adaptively chooses $L$ on-the-fly to avoid wasted computation. However, these algorithms tend to be computationally expensive, requiring many additional density, gradient, or Hessian evaluations. This will be evident when we compare Stan and NUTS to our strategy in Section~\ref{exp1}.


\section{New Sampling Methodology}

Our new MCMC algorithm for the GLPM is composed of three novel components: a split Hamiltonian Monte Carlo \citep{shahbaba2014split} integrator to update the latent positions (Section~\ref{splitHMC}), a Firefly Monte Carlo (FlyMC; \cite{maclaurin2015firefly}) auxiliary variable scheme to sub-sample non-edge dyads (Section~\ref{FlyMC}), and Gibbs sampling strategies to update the parameters $\tau$ and $\gamma^2$ of the link probability function $K$ (Section~\ref{learningK}). We present each of these contributions in sequence. 

\subsection{Split Hamiltonian Monte Carlo}\label{splitHMC}

Though it is certainly the most popular for HMC, the leapfrog integrator is just one of many options for integrating Hamiltonian dynamics \citep{leimkuhler2004simulating, chao2015exponential, mannseth2016application}. Here, we consider an alternative called \emph{split Hamiltonian Monte Carlo} \citep{shahbaba2014split}. Split HMC is a variant on the leapfrog strategy that efficiently simulates Hamiltonian motion by exploiting a Gaussian component of the posterior. It works best when the Gaussian component is a good approximation for the entire posterior. Split HMC also provides a natural choice for the mass matrix $M$ that is locally adaptive without having to evaluate the Hessian.

The standard leapfrog update described in Section~\ref{HMC} is equivalent to decomposing the energy into three terms:
\begin{equation}
H(Z,U) = -\frac{1}{2}\log(p(Z)) - \log(q(U)) - \frac{1}{2}\log(p(Z)). \label{leapdecomp}
\end{equation}
then cycling through isolated updates according (\ref{DE1}) and (\ref{DE2}) for each of the components individually. This ``split'' of the energy ensures that only one of $Z$ or $U$ is being updated at any given time, causing each isolated operation to be straightforward. In split Hamiltonian Monte Carlo, we consider a different split of the energy function, decomposing it to exploit partial analytic solutions of Hamiltonian equations.

Hamilton's equations usually lack an analytic solution, but a notable exception is when the energy is defined as the negative logarithm of a multivariate Gaussian density \citep{pakman2014exact}; the exponential and uniform distributions are others \citep{bloem2016slice}. For the Gaussian case, the motion and momentum updates can be simulated exactly along an ellipse (i.e. a contour of the Multivariate Gaussian distribution). Alone, this fact would have limited utility for Bayesian computation; exact algorithms for inference involving Gaussian posteriors are readily available. However, these analytic solutions can be remarkably useful as part of a splitting strategy.

The split Hamiltonian integrator \citep{shahbaba2014split} alternates between joint position momentum updates based on the analytical solution of the Gaussian component of the posterior and updates to the momentum to correct for the remaining portion of the posterior. When the exact part is a good approximation for the entire posterior, this allows for a coarser $\epsilon$ to maintain a high acceptance rate. We now present the decomposition of the GLPM posterior into its Gaussian and non-Gaussian components. Specifically, the likelihood of the edges, the prior density, and the momentum forming the Gaussian component, and the likelihood of non-edges forms the remainder.

Treating the $\tau$ and $\gamma^2$ as known, the posterior (\ref{kpost}) takes the form
\begin{equation}
\mathbb{P}(Z|A, \tau, \gamma^2) \propto \left(\prod_{\left(i,j\right) \in E_A}\tau_{x_{ij}}\right) \exp{\left( \frac{\sum_{\ell =1}^d Z_{\cdot \ell}^T \left(\Omega + \frac{L^A }{\gamma^2} \right) Z_{\cdot \ell}}{-2} \right)} \prod_{(i,j) \notin E_A}  \left(1-\tau_{x_{ij}} \exp{\left( \frac{\|z_i - z_j\|^2}{-2\gamma^2} \right)}\right) \label{split}
\end{equation}
where $L^A$ denotes the Laplacian of $A$. This posterior is the proportional to the product of the two components $\mathbb{P}(Z|A, \tau, \gamma^2) \propto \mathbb{P}_1(Z|A,\tau, \gamma^2) \mathbb{P}_0(Z|A, \tau, \gamma^2)$ where  
\[
\mathbb{P}_1(Z|A, \tau, \gamma^2) =   \left(\prod_{\left(i,j\right) \in E_A}\tau_{x_{ij}}\right) \exp{\left( - \frac{1}{2}\sum_{\ell =1}^d Z_{\cdot \ell}^T \left(\Omega + \frac{1}{\gamma^2} L^A \right) Z_{\cdot \ell} \right)}
\]
corresponds to the contribution of the prior and likelihood of the observed edges and  
\[
\mathbb{P}_0(Z|A, \tau, \gamma^2) =  \prod_{\left(i,j\right) \notin E_A}  \left(1-\tau_{x_{ij}} \exp{\left(-\frac{1}{2\gamma^2} \|z_i - z_j\|^2 \right)}\right)
\]
corresponds to the contribution to the likelihood of the non-edges. Using the shorthand
\begin{equation} \label{sigma}
    \Sigma = \left(\Omega + \frac{1}{\gamma^2} L^A \right), 
\end{equation}
we can now split the corresponding energy as
\begin{equation} \label{splitenergy}
    H(Z, U) =  \left[-\frac{1}{2}\log(\mathbb{P}_0(Z|A, \tau, \gamma^2))\right]  + \left[ \frac{1}{2}\sum_{\ell =1}^d \left( Z_{\cdot \ell}^T \Sigma Z_{\cdot \ell} + U_{\cdot \ell}^T M^{-1} U_{\cdot \ell}  \right) \right]  - \left[\frac{1}{2}\log(\mathbb{P}_0(Z| A, \tau, \gamma^2))\right],
\end{equation}
ignoring additive constants. The center term in this split is Gaussian. 

In the above, we have departed from the typical notation in our definition of the mass matrix $M$ and momentum variables $U$. In standard presentations of HMC (including our Section~\ref{HMC}), the target parameters and momentum variables are naturally represented as vectors. However, for a LPM, the parameters $Z$ are more suitably represented as a $n \times d$ matrix. We have thus chosen to also represent the momentum variables $U$ as a $n \times d$ matrix. Since there are $n \times d$ momentum variables, the standard notation/definition of the mass matrix would require that $M \in \mathbb{R}^{nd \times nd}$. We have opted to instead define the full mass matrix block diagonally, using $d$ repetitions of the same matrix $M \in \mathbb{R}^{n \times n}$. This choice facilitates the more compact representation in (\ref{splitenergy}) without altering the validity of the algorithm.

On top of being notationally and computationally convenient, the use of an identical mass matrix across all dimensions is justified by symmetry in the target posterior---the marginal distribution of each column of $Z$ is the same \citep{shortreed2006positional}.

The above decomposition thus suggests a natural choice of $M$. Recall from Section~\ref{HMC} that the precision matrix of the posterior is an efficient choice for $M$. Accordingly, we suggest that $\Sigma$ is a reasonable choice for $M$, as it should be a good approximation of the posterior precision matrix provided that $\mathbb{P}_0$ is a good approximation of the full posterior. Moreover, the choice $M = \Sigma$ is also particularly amenable to simulating the split HMC trajectories because it leads to arithmetic cancellations that simplify computation. Finally, the mass matrix $M$ depends on the parameter $\gamma^2$---when combined with a Monte Carlo strategy for inferring $\gamma^2$ (such as the we present in Section~\ref{learningK}), setting $M = \Sigma$ allows for the mass matrix to evolve adaptively with the state of $\gamma^2$ in the chain.

The following is a complete recipe for split HMC for LPMs, using the block diagonal mass matrix we have just defined. Note that the intermediate variable $V \in \mathbb{R}^{n \times d}$ introduced in Step 2 is a change of variable for efficiently parametrizing the contour of the multivariate Gaussian, and Step 4 inverts the change of variable to recover $U$. For more details on the exact simulation of HMC for Multivariate Gaussians, see \cite{pakman2014exact}.

Suppose that $A$ denotes an observed adjacency matrix, $\tau \in [0,1]^C$ and $\gamma^2 > 0$ denote the values of the parameters of the Gaussian link function, and $\epsilon > 0$, $L \in \mathbb{N}$ and $\Sigma$ (as defined in \ref{sigma}) denote the user-specified tuning parameters for split HMC. Given $(Z^{j}, U^{j})$, the next split HMC draw $(Z^{j+1}, U^{j+1})$ is obtained via the following steps:\\

\noindent\fbox{%
    \parbox{\textwidth}{
    \textbf{Algorithm 1: Split HMC Update}
    \begin{enumerate}
\item Update the momentum variables via Gibbs $U^{j'} \sim \textrm{MVN}(0, \Sigma)$.
\item Define intermediate variables $V^{j'} = \Sigma^{-1} U^{j'}$ and $Z^{j''} = Z^{j}$.
\item Integrate Hamiltonian motion $(Z^{j}, U^{j'}) \rightarrow (Z^{j''}, U^{j''})$ for $T = L \epsilon$ time units\\
by iterating the following updates $L$ times:
\begin{align*}
V^{j'} &\leftarrow V^{j'} + \frac{\epsilon}{2} \Sigma^{-1} \frac{\partial \log(\mathbb{P}_0(Z|A, \tau, \gamma^2))}{\partial Z} (Z^{j''})\\
(Z^{j''}, V^{j'}) &\leftarrow \left(\sin(\epsilon) V^{j'} + \cos(\epsilon) Z^{j''}, \cos(\epsilon) V^{j'} - \sin(\epsilon) Z^{j''} \right)\\
V^{j'} &\leftarrow V^{j'} + \frac{\epsilon}{2} \Sigma^{-1} \frac{\partial \log(\mathbb{P}_0(Z|A, \tau, \gamma^2))}{\partial Z} (Z^{j''})
\end{align*}
Finish by setting $U^{j''} = \Sigma V^{j'}$.
\item Accept the move $(Z^{j+1}, U^{j+1}) = (Z^{j''}, -U^{j''})$ with probability
\[
\textrm{min}\left( \frac{\mathbb{P}(Z^{j''}|A, \tau, \gamma^2)}{\mathbb{P}(Z^{j}|A, \tau, \gamma^2)}   \exp\left(\frac{1}{2}\sum_{\ell =1}^d \left(U'_{\cdot \ell} - U''_{\cdot \ell}\right)^T \Sigma^{-1} \left(U'_{\cdot \ell} - U''_{\cdot \ell}  \right)\right),  1\right).
\]
Otherwise, the move is rejected and $(Z^{j+1}, U^{j+1}) = (Z^{j}, U^{j'})$. 
\end{enumerate}
}%
}
\\

In Step 3 of Algorithm 1, the gradient functions return $n \times d$ matrices defined by
\begin{align} 
 \left(\frac{\partial \log(\mathbb{P}_0(Z|A, \tau, \gamma^2))}{\partial Z}(Z)\right)_{ik} &=  \frac{\partial \log(\mathbb{P}_0(Z|A, \tau, \gamma^2))}{\partial Z_{ik}}\\
 &= \sum_{j: (i,j) \notin E_A} \frac{\left(Z_{ik} - Z_{jk}\right)}{\gamma^2}\frac{\tau_{x_{ij}} \exp\left(-\frac{\|z_i - z_j \|^2}{2 \gamma^2}\right)}{1 - \tau_{x_{ij}}\exp\left(-\frac{\|z_i - z_j \|^2}{2 \gamma^2}\right)}. \label{gradient}
\end{align}

\subsection{Firefly Sampling of Non-Edges} \label{FlyMC}

Recall from Section~\ref{splitHMC} that exact simulation of the Hamiltonian motion is thwarted by the non-edge terms in the likelihood. Furthermore, the computational bottlenecks for running Algorithm 1 are the gradient evaluations in Step 3 and the acceptance ratio evaluation in Step 4. These computations each require an operation to be performed for each non-edge, making them especially expensive for large sparse networks due to the large number of non-edges. It can thus be beneficial to eliminate some non-edge terms from the likelihood at each iteration of split HMC. Here, we propose such a strategy.

Consider the following data augmentation scheme inspired by the Firefly Monte Carlo (FlyMC; \cite{maclaurin2015firefly}). For each $(i,j) \in [n]^2$, we define auxiliary independent binary random variables $\theta_{ij}$ such that $\mathbb{P}(\theta_{ij} = 1| \tau_{x_{ij}}) = \tau_{x_{ij}}$. Using these auxiliary variables, we can re-express the edge probabilities as 
\begin{align*}
\mathbb{P}(A_{ij} = 1 | \theta_{ij} = 1, \tau_{x_{ij}}, \gamma^2, z_i, z_j) &=  \exp{\left(-\frac{1}{2\gamma^2} \|z_i - z_j\|^2 \right)},\\
\mathbb{P}(A_{ij} = 1 | \theta_{ij} = 0, \tau_{x_{ij}}, \gamma^2, z_i, z_j) &=  0,
\end{align*}
while maintaining the same marginal likelihood. Now,
\begin{align*}
\mathbb{P}(\theta_{ij} = 0 | A_{ij} = 0, Z, \tau_{x_{ij}}, \gamma^2) &= \frac{1 - \tau_{x_{ij}}}{1 - \tau_{x_{ij}} \exp{\left(-\frac{1}{2\gamma^2} \|z_i - z_j\|^2 \right) }},\\
\mathbb{P}(\theta_{ij} = 0 | A_{ij} = 1, Z, \tau_{x_{ij}}, \gamma^2) &= 0.
\end{align*}

Note that for all $(i,j) \in E_A$, $\theta_{ij} = 1$ must hold. Thus, 
\[
\mathbb{P}(Z|A, \theta, \tau, \gamma^2) = \mathbb{P}_1(Z|A, \tau, \gamma^2) \prod_{ij: \theta_{ij} = 1, A_{ij} = 0}  \left(1- \exp{\left(-\frac{1}{2\gamma^2} \|z_i - z_j\|^2 \right)}\right), 
\]
meaning that 
\[
\mathbb{P}_0^*(Z|A, \theta, \tau, \gamma^2) = \prod_{ij: \theta_{ij} = 1, A_{ij} = 0}  \left(1- \exp{\left(-\frac{1}{2\gamma^2} \|z_i - z_j\|^2 \right)}\right)
\]
can replace $\mathbb{P}_0(Z|A, \tau, \gamma^2)$ in Split HMC once the $\theta$ variables are instantiated. If many of the $\theta_{ij}$ are 0, computing $\mathbb{P}_0(Z|A, \tau, \gamma^2)$---and its gradients---is far cheaper than computing the marginal $\mathbb{P}_0(Z|A, \tau, \gamma^2)$ analogs. Combining this data augmentation strategy with split HMC can be a major computational improvement, provided that we can update the $\theta_{ij}$ values efficiently.

To do so, we propose a Metropolis-Hastings step using proposal $q(\theta_{ij} = 1) = \tau_{ij}$. Let $\textrm{MH}(\theta_{ij} = 0 \rightarrow \theta_{ij} = 1)$ denote the Metropolis-Hastings ratio associated with a proposed move from $\theta_{ij} = 0$ to $\theta_{ij} = 1$, and let $\textrm{MH}(\theta_{ij} = 1 \rightarrow \theta_{ij} = 0)$ denote the Metropolis-Hastings ratio associated with a proposed move from $\theta_{ij} = 1$ to $\theta_{ij} = 0$. The values of these ratios are given by
\begin{align}
\textrm{MH}(\theta_{ij} = 0 \rightarrow \theta_{ij} = 1) &=  \left(1- \exp{\left(-\frac{1}{2\gamma_{ij}^2} \|z_i - z_j\|^2 \right)}\right),\\
\textrm{MH}(\theta_{ij} = 1 \rightarrow \theta_{ij} = 0) &=  \frac{1}{1- \exp{\left(-\frac{1}{2\gamma_{ij}^2} \|z_i - z_j\|^2 \right)}} \geq 1.
\end{align}
Thus, out of the four possible moves $0 \rightarrow 0$, $0 \rightarrow 1$, $1 \rightarrow 0$, $1 \rightarrow 1$, the accept reject step need only be performed for $0 \rightarrow 1$.  In contrast, a Gibbs update of $\theta_{ij}$ from its full conditional would require that the posterior density be evaluated for any of the four moves. As such, this Metropolis update involves less computation than a full Gibbs update, especially for small values of $\tau$.

Going forward, we refer to the parameter augmentation and update strategy described above as FlyMC. The reduction in computational cost of evaluating the posterior density and gradients under FlyMC is most prevalent when most of the $\theta_{ij}$ are 0. Because $\mathbb{P}(\theta_{ij} =0| \tau_{x_{ij}}) = 1 - \tau_{x_{ij}}$, the computational gains from FlyMC are largest when $\tau_{x_{ij}}$ is small. On the other hand, when $\tau_{x_{ij}}$ are relatively large (close to 1), most values of the $\theta_{ij}$ will be one, meaning the computational improvements in evaluating the gradient may not justify the computational expense of instantiating and updating the $\theta$ variables. In the extreme case of $\tau_{x_{ij}} = 1$, no subsampling will occur at all, so FlyMC should not be included.

For sparse networks, however, $\tau_{x_{ij}}$ may be very small for some values of $x_{ij}$, leading to substantial computational gains. In addition to providing a computational speed-up, the new FlyMC posterior facilitates the inference of $\tau$ via Gibbs steps.

\subsection{Bayesian Inference of the Parameters of the Link Function} \label{learningK}

Thus far, our posterior computation strategy for $Z$ has held $\tau$ and $\gamma^2$---the parameters of the link function---at fixed values. In most applications, $\tau$ and $\gamma^2$ are unknown---they need to be inferred along with $Z$. As we noted in Section~\ref{LPMBayes}, $\tau$ may even be the primary inferential target. It is thus important that our posterior computation strategy compute the full joint posterior of $\tau$, $\gamma^2$, and $Z$. Here, we describe efficient Gibbs updates for both $\tau$ (Section~\ref{tau}) and $\gamma^2$ (Section~\ref{gamma2}) to be alternated with our split HMC + FlyMC.

The updates we describe here apply to specific families of priors on $\tau$ and $\gamma^2$. In particular, we use independent Beta priors for each entry in $\tau$, along with an inverse Gamma prior $\gamma^2$. Moreover, the update for $\tau$ is applicable only in conjunction with the FlyMC strategy outlined in Section~\ref{FlyMC}. If FlyMC is not used, we recommend a simple random walk Metropolis-Hastings update for $\tau$ instead. Finally, our update strategy for $\gamma^2$ depends on the centered re-parameterization mentioned in Section~\ref{lpmdefinition} where $\gamma$ is treated as a scaling factor for the latent positions. It is applicable whether or not FlyMC is used. An detailed expression of the full posterior and relevant conditionals are available in Section~\ref{math} of the Appendix.

\subsubsection{Updating $\tau$ given $Z, \theta, \gamma^2$} \label{tau}

We assume independent $\textrm{Beta}(\alpha_c, \beta_c)$ priors on the entries in $\tau$, with $c \in [C]$ indexing the possible levels of the covariate, and $\alpha_c, \beta_c \in \mathbb{R}_+$. If Section~\ref{FlyMC}'s FlyMC strategy is used, each $\tau_c$ can then be updated according to its conditional posterior distribution given the FlyMC variables $\theta$ and the covariates $x$. Indeed, this posterior distribution is conjugate to the Beta prior because inferring $\tau_c$ given $\theta$, and the covariate values $x$ is equivalent to inferring the probability parameter of a sequence of Bernoulli trials (specifically the $\theta_{ij}$ for which $x_{ij} = c$). Thus,
\begin{equation}
\tau_c \mid \theta,x \sim \textrm{Beta}(\alpha_c + \Theta^1_c, \beta_c + \Theta^0_c), \label{conditionaltau}
\end{equation}
where $\Theta^0, \Theta^1 \in (\left\{0\right\} \cup \mathbb{N})^C$ are defined according to
\begin{align*}
\Theta^0_c &= |\left\{ (i,j) \in [n]^2: \theta_{ij} = 0 \textrm{ and } x_{ij} = c \right\}| \\
\Theta^1_c &= |\left\{ (i,j) \in [n]^2: \theta_{ij} = 1 \textrm{ and } x_{ij} = c \right\}| 
\end{align*}
This update can be made efficient by keeping track of $\Theta^0$ and $\Theta^1$ during FlyMC updates.

\subsubsection{Updating $\gamma^2$ given $Z$, $\tau$, $\theta$} \label{gamma2}

We assume a InverseGamma($a$, $b$) prior on $\gamma^2$, where $a,b > 0$. Then, the posterior density of $\gamma^2 | A, Z, \theta$ is proportional to
\[
p(\gamma^2 \mid A, Z, \theta) \propto \textrm{IG}(\gamma^2| a,b) \exp{\left(  \frac{-\sum_{\ell =1}^d Z_{\cdot \ell}^T  L^A Z_{\cdot \ell}}{2\gamma^2} \right)} \prod_{ij: \theta_{ij} = 1, A_{ij} = 0}  \left(1- \exp{\left(-\frac{\|z_i - z_j\|^2}{2\gamma^2}  \right)}\right)
\]
where IG$(\gamma^2|a,b)$ denotes the probability density function of the InverseGamma($a$, $b$) distribution. This density provides no closed-form Gibbs update and is expensive to evaluate. 

Alternatively, we propose using the centered parameterization (\ref{breakout}) in Section~\ref{lpmdefinition}. By modifying the prior on $Z$ to depend on $\gamma$ through $Z_{\cdot \ell} \sim N(0, \gamma^{-2} \Omega)$ for $\ell \in [d]$, we can instead treat $\gamma^2$ as known to be 1 in the link function (\ref{gaussianlink}). Instead, we infer a scale parameter for the variance of the latent positions.

Accordingly, the conditional distribution of $\gamma^2$ given $A, Z, \theta, \tau, x$ is given by
\begin{align*}
p(\gamma^2 \mid A, Z, \theta, x) &= \textrm{IG}(\gamma^2| a,b) \exp{\left( - \frac{\gamma^2}{2}\sum_{\ell =1}^d \left(Z_{\cdot \ell}\right)^T  \Omega^{-1} Z_{\cdot \ell} \right)}\\
                                                              & \propto \textrm{IG}\left(\gamma^2| a + \frac{nd}{2}, b + \frac{1}{2} \sum_{\ell =1}^d \left(Z_{\cdot \ell}\right)^T  \Omega^{-1} Z_{\cdot \ell}\right).
\end{align*}
The posterior dependence of $\gamma^2$ on $\theta$ and $A$ has been eliminated and the prior on $\gamma^2$ is now conjugate, allowing for a simple Gibbs update. 

This reparameterization and the corresponding Gibbs update is straightforward to incorporate with the other updates of $Z$, $\theta$, and $\tau$ described in Sections~\ref{splitHMC}, Section~\ref{FlyMC}, and Section~\ref{tau}, respectively. We simply fix the $\gamma^2$ in the link function at 1, and replace $\Omega$ with $\gamma^{-2} \Omega$ for the prior covariance in those sections. After performing MCMC sampling in this re-parametrized setting, one can recover the original parameterization with $\gamma^2$ in the link function by scaling the $Z$ draws.

Sections~\ref{splitHMC}, \ref{FlyMC}, and \ref{learningK} a full MCMC strategy for posterior computation of $Z, \tau, \gamma^2$ that also introduces and updates auxiliary FlyMC variables $\theta$. Each sweep of the chain iterates through a split HMC update of $Z$, a Metropolis update of each $\theta$, a Gibbs update for each entry in $\tau$, and a Gibbs update of $\gamma^2$. Alternatively, if FlyMC is not incorporated, posterior computation of $Z$, $\tau$, and $\gamma^2$ can be performed by alternating the split HMC update of $Z$, a Metropolis update of each entry in $\tau$, then a Gibbs update of $\gamma^2$. For completeness, the functional form of the posterior being computed in both the split HMC and split HMC +FlyMC algorithms is outlined in Section~\ref{math} of the Appendix.

\section{Empirical Studies}\label{experiments}

To explore and understand the relative strengths and weaknesses of our new posterior inference algorithm for GLPMs, we conduct two empirical studies. Study 1 consists of a ``bake-off'' between various inference algorithms, comparing the efficiency of our method to that of plausible competitors from in literature. These comparisons span a variety of synthetically generated networks of different sizes and sparsity levels. Study 2 is a real data example, demonstrating efficiency of split HMC and split HMC + FlyMC compared to traditional Metropolis within Gibbs for modeling information-sharing among elementary school teachers and staff in a school district. Multiple model set-ups are considered in Study 2, including the use of use categorical covariates and longitudinal network data. 

Given our interest in unbiased methods for posterior computation, we focus exclusively on MCMC algorithms in both experiments. These MCMC algorithms all approximate posterior distributions equally well if run for a sufficiently long time, so our comparisons are based on their relative efficiency. Descriptions of the software and hardware we use for the experiments, details of how we tune the algorithms, and details of how we compare their performance are available in Sections~\ref{compdetails}, \ref{tuning}, and \ref{ESS} of the Appendix, respectively.

\subsection{Study 1: Synthetic Data} \label{exp1}

In this empirical study, we investigate the efficiency of split HMC (Section~\ref{splitHMC}) and split HMC + FlyMC (Sections~\ref{splitHMC} and \ref{FlyMC}) compared to nine other exact MCMC algorithms from the literature. We are especially interested in fitting LPMs for large sparse networks, so we have tooled the study design to compare performance as networks get larger and more sparse. Our investigation involves fitting Gaussian LPMs to sixteen different synthetically generated networks. These networks---stochastically generated according to a GLPMs with pre-specified values of $\tau$, $\gamma^2$, and $n$---demonstrate a variety of different sizes and sparsity. Specifically, we consider the full factorial design outlined by the first three columns of Table~\ref{design1}. No covariates are included, and all latent positions are drawn from a two-dimensional isotropic Gaussian. Note that the type of sparsity driven by small $\tau$ has different structure than that driven by small $\gamma^2$, so our design considers both.

\begin{figure}[!h] 
\centering
\resizebox{\textwidth}{8cm}{
\input{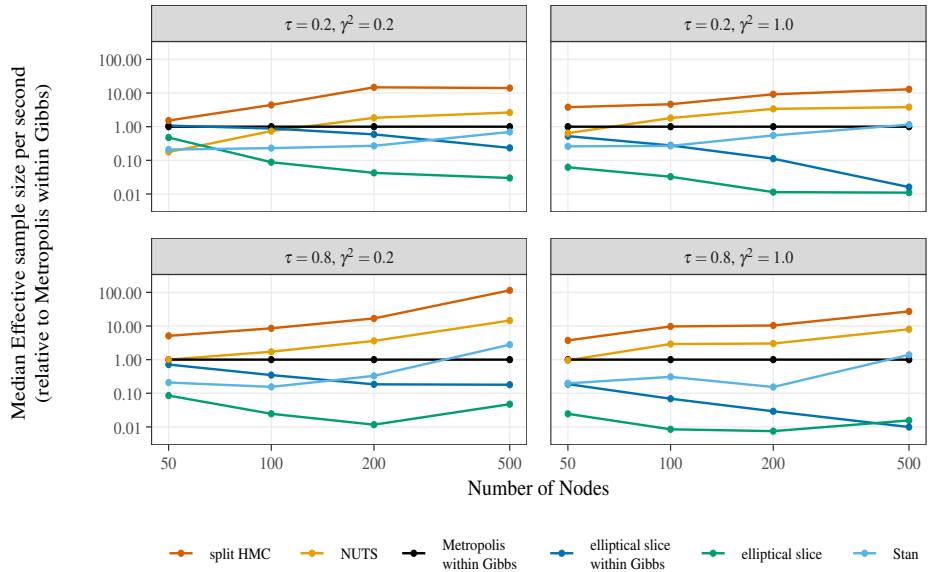}
}
\caption{A depiction of the relationship between the number of nodes in the synthetically generated networks ($\tau = 0.2, 0.8, \gamma^2 = 0.2, 1.0$) for Empirical Study 1 and the relative efficiency of the five posterior computation algorithms. For each algorithm, relative efficiency (vertical axis) is quantified as the median across 500 dyads in the synthetic network of the relative Markov chain efficiency compared to Metropolis within Gibbs as a baseline. Note that the vertical axis is on the log scale.} \label{experiment1afig} 
\end{figure}

\begin{figure}[!h] 
\centering
\resizebox{\textwidth}{8cm}{
\input{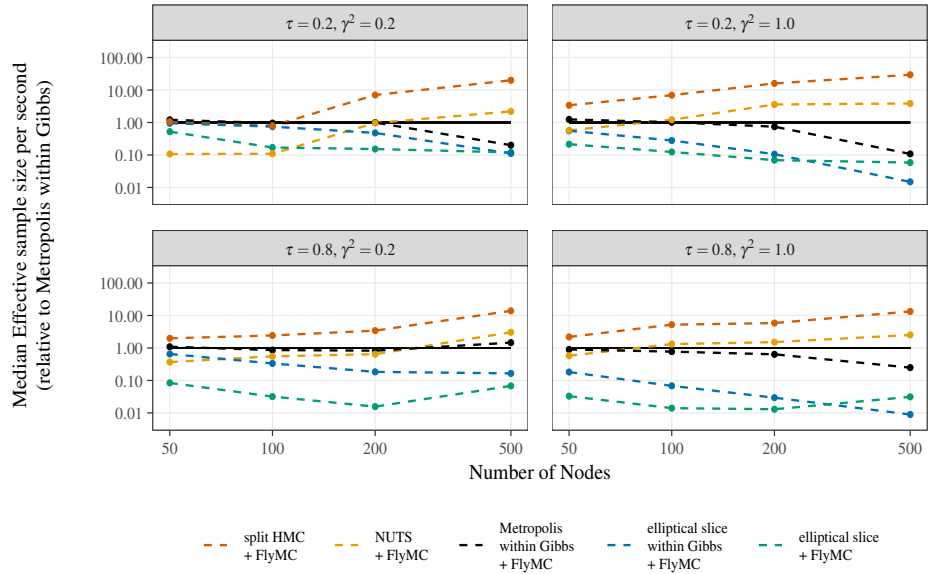}
}
\caption{A depiction of the relationship between the number of nodes in the synthetically generated networks ($\tau = 0.2, 0.8, \gamma^2 = 0.2, 1.0$) for Empirical Study 1 and the relative efficiency (compared to Metropolis within Gibbs) of the five FlyMC posterior computation algorithms. The solid black baseline is included for easy comparison to Metropolis within Gibbs. Note that the vertical axis is on the log scale.} \label{experiment1bfig} 
\end{figure}

Figures~\ref{experiment1afig} and \ref{experiment1bfig} report the relative performance of eleven different MCMC posterior inference algorithms across the 16 networks described in Table~\ref{design1}. The algorithms vary along two criteria: the proposal used to update $Z$, and whether or not FlyMC (Section~\ref{FlyMC}) is used to subsample the non-edges. We consider five strategies for updating $Z$: Metropolis within Gibbs (Section~\ref{LPMBayes}), elliptical slice sampling \citep{murray2010elliptical}, elliptical slice sampling within Gibbs \citep{hahn2019efficient}, split HMC (Section~\ref{splitHMC}) with $T \approx 2$, and an alternative implementation of Split HMC that uses NUTS \citep{hoffman2014no} to adaptively choose the integration time $T$. For each of these strategies we consider implementations both with and without FlyMC. Finally, we include HMC as implemented Stan version 2.18.2 \citep{carpenter2017stan} as an additional competitor algorithm, bringing the total number of algorithms to eleven. A FlyMC version of Stan is not possible. Metropolis within Gibbs serves as a standard baseline, the elliptical slice sampling algorithms represent alternative ways to exploit the Gaussian component in the posterior, the NUTS version of Split HMC illustrates the extra computational cost of adaptively choosing $T$, and Stan represents existing software.

In addition to sampling the latent positions, we use each algorithm to sample $\tau$ using a uniform prior and $\gamma^2$ using an inverse gamma $\textrm{IG(1,1)}$ prior. For the five FlyMC algorithms, we alternate between updates of $Z$, updates of the FlyMC variables $\theta$ according to the Metropolis strategy outlined in Section~\ref{FlyMC}, updates of $\tau$ according to the Gibbs strategy outlined in Section~\ref{tau}, and an update of $\gamma^2$ to the Gibbs strategy outlined in Section~\ref{gamma2}. Where FlyMC is not used, $\tau$ is updated using a random walk Metropolis algorithm instead of Gibbs. 

Each algorithm was initialized identically using the maximum likelihood estimate of $Z$ and the true values of $\tau$ and $\gamma^2$. The FlyMC and non-FlyMC versions of each algorithm were each simulated for 10000 iterations. The Stan chain was run for just 2000 iterations due to its much longer runtime. The various hyperparameters (step size $\epsilon$ for HMC and random walk standard deviation $\delta$ for Metropolis) are shown in Table~\ref{design1}; these values were tuned according to the strategies outlined in Section~\ref{tuning} of the Appendix. 

To evaluate the relative performances of each algorithm on each network, we consider the effective number of samples generated per second of runtime. The effective sample size $ESS_f$ of a Markov chain $\theta^1, \theta^2, \ldots, \theta^N$ is defined as 
\[
\textrm{ESS}_{f}(\theta^1,\ldots, \theta^N) = \frac{N}{1 + 2 \sum_{t=1}^{\infty} \rho_{t,f}}
\]
where $\rho_{t,f}$ denotes the $t$-lag autocorrelation of the function $f(\theta)$ in the chain. For each network, we compute the $ESS_f$ of each algorithm for 500 distinct choices of $f$. Each choice corresponds to the log probability of an edge at a randomly selected dyad. For each dyad, we calculate
\begin{equation} \label{ESS2}
\frac{\textrm{ESS}_{f}(\theta^1,\ldots, \theta^N)}{\textrm{ESS}_{f}(\theta_m^1,\ldots, \theta_m^N)} \times \frac{\textrm{time (in seconds) taken to compute the chain }  \theta_m^1,\ldots, \theta_m^N}{\textrm{time (in seconds) taken to compute the chain } \theta^1,\ldots, \theta^N}
\end{equation}
where $\theta^1, \theta^2, \ldots, \theta^N$ denotes draws according to the algorithm being evaluated, $\theta_m^1, \ldots, \theta^N_m$ denotes draws according to a well-tuned Metropolis within Gibbs algorithm exploring the same posterior, and $f$ denotes the log probability of the given dyad. Figure~\ref{experiment1afig} reports the median of (\ref{ESS2}) across 500 dyads for each of the non-FlyMC algorithms. For readability, the results from the analogous FlyMC algorithms are presented separately as Figure~\ref{experiment1bfig}, using the same colors (but dashed instead of solid lines). For more explanation and justification of the evaluation metric in (\ref{ESS2}), see Section~\ref{ESS} of the Appendix. 

\setlength{\tabcolsep}{9pt}
\begin{table}
  \caption{A summary of the simulation design for Study 1. The first three columns specify the number of nodes, the link function parameter $\gamma^2$, and the link function parameter $\tau$ used to generate each of the synthetic networks. The edge density column reports the fraction of dyads exhibiting edges in each network. The final two columns report the tuned values of hyperparameters for the algorithms.}
  \label{design1}
    \begin{tabular}{l c c c c  c}
      \hline
      \makecell{Number\\ of Nodes} & $\tau$ &
      $\gamma^2$&
      \makecell{ Edge \\ Density } & \makecell{Step size\\ $\epsilon$ for HMC\\ (+ FlyMC)}  &
      \makecell{Step size $\delta$\\ for Metropolis\\ (+ FlyMC)} \\
      \hline
50 & 0.20 & 0.20 &  0.022&0.14 (0.08) & 2.40 (2.40) \\ 
50 & 0.80 & 0.20 & 0.087&0.29 (0.16) & 1.20 (1.00)\\ 
50 & 0.20 & 1.00 & 0.079&0.57 (0.21) & 1.20 (1.20) \\ 
50 & 0.80 & 1.00 & 0.304&0.42 (0.28) & 0.60 (0.60) \\ 
100 & 0.20 & 0.20 & 0.022&0.20 (0.06) & 2.40 (1.60) \\ 
100 & 0.80 & 0.20 & 0.083&0.23 (0.15) & 0.80 (0.80) \\ 
100 & 0.20 & 1.00 & 0.075&0.40 (0.20) & 1.00 (0.80) \\ 
100 & 0.80 & 1.00 & 0.285&0.31 (0.28) & 0.40 (0.40) \\ 
200 & 0.20 & 0.20 & 0.02&0.27 (0.08) & 1.60 (1.20) \\ 
200 & 0.80 & 0.20 & 0.08&0.25 (0.18) & 0.60 (0.60) \\ 
 200 & 0.20 & 1.00 & 0.07&0.28 (0.19) & 0.80 (0.60) \\ 
200 & 0.80 & 1.00 & 0.28&0.28 (0.25) & 0.30 (0.30) \\ 
500 & 0.20 & 0.20 & 0.016&0.27 (0.08) & 0.80 (0.80) \\ 
500 & 0.80 & 0.20 & 0.066&0.20 (0.13) & 0.40 (0.40) \\ 
500 & 0.20 & 1.00 & 0.062&0.18 (0.18) & 0.50 (0.40) \\ 
500 & 0.80 & 1.00 & 0.254&0.16 (0.17) & 0.20 (0.20) \\ 
      \hline
    \end{tabular}
  \end{table}

The results shown in Figure~\ref{experiment1afig} and Figure~\ref{experiment1bfig} demonstrate several phenomena. Notably, Split HMC and and Split HMC + FlyMC are the standout performers across all networks considered. Split HMC clearly outperforms Metropolis within Gibbs for all networks, and Split HMC + FlyMC outperforms Metropolis within Gibbs for all networks except the smaller networks in the sparsest regime. Notably, both implementations of Split HMC with $T\approx 2$ outperform their NUTS counterparts and Stan, demonstrating that the extra computational cost of using NUTS or Stan to adaptively updating $L$ may be unwarranted for LPMs. 

All methods based on elliptical slice sampling perform poorly, demonstrating that HMC is a better method for exploiting the near-Gaussianity of the posterior than elliptical slice sampling. Indeed, the elliptical slice sampling algorithms performed worse than Metropolis within Gibbs. The poor performance of the elliptical slice within Gibbs algorithms was due to its runtime---the  conditional means and variances of each latent position at each iteration are very expensive to compute. The joint update elliptical slice algorithms performed poorly for the opposite reason. They had much faster runtimes, but the corresponding chains mixed very slowly because the draws exhibited very high autocorrelation.

The dominance of the Split HMC methods appears to be more pronounced for larger networks. For the denser $\tau = 0.8$ networks, Split HMC performs remarkably well, showing a distinct upward trend, indicating that its dominance over Metropolis within Gibbs would be even more pronounced for larger networks. For the sparser $\tau = 0.2$ networks, Split HMC + FlyMC demonstrates a similar upward trend. The tuned values of $\epsilon$ and $\delta$ shown in Table~\ref{design1} demonstrate that split HMC is more robust to large network sizes: while the tuned step sizes for the Metropolis within Gibbs decay as the number of nodes increases, the tuned values of $\epsilon$ for split HMC remain more stable.

\begin{figure}[!h] 
\centering
\resizebox{\textwidth}{8cm}{
\input{Experiment1cnew.tex}
}
\caption{Boxplots showing the relative efficiency of Split HMC + FlyMC and Split HMC relative to Metropolis within Gibbs across 500 dyads in each network.} \label{experiment1cfig} 
\end{figure}

To facilitate comparison between the FlyMC and non-FlyMC implementations of split HMC, we have also included Figure~\ref{experiment1cfig}. Instead of just reporting the median, it summarizes the entire distribution of (\ref{ESS2}) across the 500 sampled dyads. From the side-by-side boxplots, we can see that split HMC clearly outperforms split HMC + FlyMC for the $\tau = 0.8$ networks of all sizes. For the $\tau = 0.2$ networks, the comparison is less clear cut. For the very sparse network $\tau = 0.2, \gamma^2 = 0.2$, split HMC outperforms the FlyMC version for the smaller networks, but FlyMC edges it out for the 500 node network. It is worth noting that in this very sparse regime for smaller networks, the extreme lack of edges can lead to ambiguity in whether the extreme sparsity is driven by small $\tau$, small $\gamma^2$, or both. The joint posteriors of $\tau, \gamma^2$ for different synthetic networks shown in Figure~\ref{tauvsgamma} in Section~\ref{adfigs} of the Appendix demonstrates this phenomenon---there is a remarkable amount of uncertainty in the posterior of $\tau$ for the smaller sparse networks.

 We have noticed that the FlyMC updates of $\tau$ and $\theta$ tend to mix slowly in these uncertain situations, thus leading to slow exploration of the joint distribution of $(\tau, \gamma^2)$. Unencumbered by slow-mixing $\theta$ variables, split HMC tends to perform better in these under-identified settings, suggesting that FlyMC should only be used when the $\tau$ variable is better identified. This seems to be the case in the $\tau = 0.2, \gamma^2 = 1.0$ networks, where the FlyMC clearly outperforms the non-FlyMC version. 

The full distribution of the relative efficiencies highlights another observation---although the split HMC algorithms tend to outperform Metropolis within Gibbs for the vast majority of dyads, there is often a small minority of dyads for which Metropolis within Gibbs performs better (seen as the lower tails of the boxplots in Figure~\ref{experiment1cfig} sometimes extending below 1). A thorough investigation into these dyads revealed no apparent pattern for which dyads tend to perform relatively poorly in a given network, suggesting that the primary explanation is simply the high-dimensionality of the posterior---with so many dimensions along which to mix, there will often be a small minority that mix more slowly. Regardless, the underperformance is not drastic for the large networks in which we are interested---split HMC still performs at the same order of magnitude as Metropolis within Gibbs.

From this empirical study, we have demonstrated that split HMC and split HMC + FlyMC tend to outperform competitors in the literature on synthetically generated data. For denser networks, or networks for which $\tau$ and $\gamma^2$ are poorly identified (i.e. smaller sparse networks), split HMC tends to be the better choice. For larger sparse networks, split HMC + FlyMC seems to be the top performer. In all cases, these strategies perform far better than simple Metropolis within Gibbs.

\subsection{Study 2: Network of Information-sharing in a School District} \label{exp2}

To demonstrate the efficacy of our split Hamiltonian Monte Carlo strategies on real data, we now showcase several applications of LPMs to information-sharing networks of teachers and staff in a school district. These applications involve several model/network configurations commonly encountered in practice: networks with covariates encoding group memberships of the nodes, longitudinally-observed networks with models promoting serial dependence of the latent positions, and combinations of the two. The data we use were collected in a mid-sized suburban school district in the Midwestern United States as part of the Distributed Leadership Studies at Northwestern University, a comprehensive program of research involving several longitudinal studies of workplace and social interactions among school staff and school systems. For more details about this particular dataset, see \citet{spillane2013organizing}; \citet{spillane2018school}.

In five separate years, elementary school teachers and staff within this district were surveyed about who in the district they went to for advice, as well as the school in which they worked and other relevant covariates. Over these years, 661 distinct individuals responded to the survey in at least one year. 129 of them were present for all five surveys.

For the purposes of this empirical study, we have compiled the survey responses into a series of five undirected \emph{information-sharing} networks---one for each year of data. These undirected information-sharing relationships were obtained by symmetrizing the information in the advice-seeking survey. That is, for each network, an edge is present between two individuals if either of them reported going to the other for advice in that year. In addition to the edge information for each dyad $(i,j)$, we have access to indicators $x_{ij1}$ and $x_{ij2}$ defined as:
\begin{align*}
x_{ij1} &= \begin{cases} 
      1 & \text{ if individuals $i$ and $j$ work at the same school} \\
      0 & \text{ otherwise}\\
   \end{cases}\\
   x_{ij2} &= \begin{cases} 
      1 & \text{ if individuals $i$ and $j$ shared advice in the previous year} \\
      0 & \text{ otherwise}\\
   \end{cases}
\end{align*}
which can be used to form covariates. From this sequence of five networks, we have extracted four different datasets, summarized in Table~\ref{dataconfigs}.

\setlength{\tabcolsep}{6pt}
\begin{table}[h]
\caption{A summary of the data configurations for Study 2} \label{dataconfigs}
  \label{design2}
    \begin{tabular}{c c c c c }
      \hline
      Dataset & \makecell{Survey Years} & \makecell{School IDs} & \makecell{Number of Nodes} & \makecell{Number of Edges} \\
      \hline
      \makecell{one year, \\ one school}&   1 &  ID 4 &   32&   150 \\ \hline
      \makecell{one year, \\ all schools}&   1 &   ID 1 to ID 14&  326&   1363 \\ \hline
      \makecell{all years, \\ one school}&         1-5 &  ID 4&  14 &  79 \\ \hline
      \makecell{all years, \\ all schools} &  1-5 &  ID 1 to ID 14 &    129 &   1038  \\ \hline
    \end{tabular}
  \end{table}

Using the datasets in Table~\ref{design2}, we fit six different models: one model for each of the \emph{one school, one year} and \emph{one school, all years} networks, and two models with different covariate configurations for each of the \emph{all schools, one year} and \emph{all schools, all years} networks. Table~\ref{design2} summarizes these six models, their covariate formulations, and their resultant model fits. For all models, we assume a uniform prior on the entries in $\tau$, a $\textrm{IG(1,1)}$ prior on $\gamma^2$, and a latent dimension $d=2$ for the latent positions. For models 1-3, we assume  independent isotropic Gaussian priors on the latent positions. Models 4-6 follow the same nodes across multiple years, so we use an autoregressive Gaussian prior; each nodes' sequence of latent positions is assumed to have an autocorrelation of 0.95 a priori, and distinct nodes are assumed to be independent.

\begin{figure}[!h] 
\centering
\resizebox{\textwidth}{8cm}{
\input{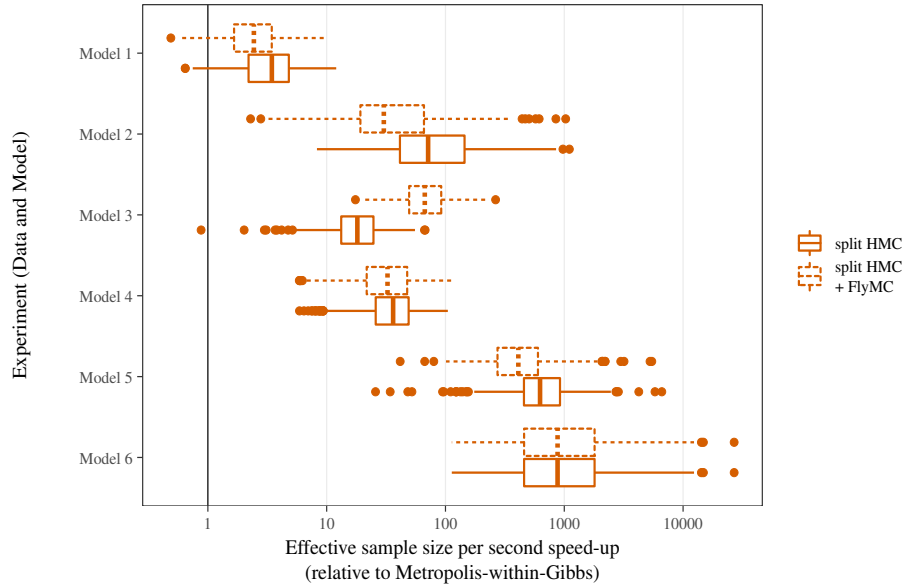}
}
\caption{Boxplots depicting the relative efficiency (in terms of effective sample size per second) of split Hamiltonian Monte Carlo (both with and without FlyMC) compared to Metropolis within Gibbs. The six different data/model configuration described above are considered. For each configuration, the relative speed-up in effective sample size per second is provided for computing the posterior log probability of a random subset of dyads in the network. Note that the horizontal axis is provided on the log scale.} \label{experiment2fig} 
\end{figure}

For all six models, we employed the same tuning strategy as in Study 1 (described in Section~\ref{tuning} of the Appendix) to choose the hyperparameters and random walk step sizes. Each Metropolis within Gibbs chain was run for 20000 iterations, and each split HMC and split HMC + FlyMC chain was run for 10000 iterations. For each algorithm and model, the relative speed-up is summarized using a boxplot for at most 500 randomly selected dyads in Figure~\ref{experiment2fig}.

    \setlength{\tabcolsep}{6pt}
\begin{table}[h]
\caption{A summary of the six model configurations and fits for Study 2. For each model, the Dataset column reports the network data which the model is fit, the $x_{ij1}$ and $x_{ij2}$ columns report the indicators used to form the link function covariate, and the $\hat{\gamma}_{\text{EAP}}^2$ and $\hat{\tau}_{\text{EAP}}$ columns report the expectations a posteriori of link function parameters $\gamma^2$ and $\tau$, respectively. Note that a different value of $\tau$ is fit for each level of the covariate.}
  \label{design3}
    \begin{tabular}{l c c c c c l}
      \hline
      Model & Dataset & \makecell{$x_{ij1} $\\ (same school)} & \makecell{$x_{ij2}$\\ (previous edge)} & $\hat{\gamma}_{\text{EAP}}^2$ & $\hat{\tau}_{\text{EAP}}$ \\
      \hline
      1&\makecell{one year, \\ one school}& Excluded & Excluded  &  1.22 & 0.85 & \\ \hline
      2 & \makecell{one year, \\ all schools}& Excluded & Excluded  & 0.15 &   0.43 & \\ \hline
      3 & \makecell{one year, \\ all schools}& Included & Excluded  & 1.18&  \makecell{ 0.86\\ 0.01} & \makecell[l]{$x_{ij1} = 1$ \\ $x_{ij1} = 0$}  \\ \hline
      4 &\makecell{all years, \\ one school}&  Excluded & Included &  1.00 &   \makecell{ 0.96\\ 0.47} & \makecell[l]{$x_{ij2} = 1$ \\ $x_{ij2} = 0$}  \\ \hline
      5 & \makecell{all years, \\ all schools} &  Excluded & Included & 0.18&  \makecell{ 0.95\\ 0.27} & \makecell[l]{$x_{ij2} = 1$ \\ $x_{ij2} = 0$}  \\ \hline      
      6 &\makecell{all years, \\ all schools}&   Included & Included & 1.09 &    \makecell{ 0.86  \\ 0.74\\0.32\\0.01} & \makecell[l]{$x_{ij1} = 1, \; x_{ij2} = 1$ \\ $x_{ij1} = 1, \; x_{ij2} = 0$\\ $x_{ij1} = 0, \; x_{ij2} = 1$\\ $x_{ij1} = 0, \; x_{ij2} = 0$}  \\ \hline 
          \end{tabular}
  \end{table}

For each of the six models, both the non-FlyMC and FlyMC versions of split HMC clearly outperform Metropolis within Gibbs, even more so than in Study 1. The speed-up is most pronounced in Model 6, with the algorithms being almost 1000 times more efficient than Metropolis within Gibbs. The two HMC algorithms perform comparably across the different models. The most noticeable difference is for models 2 and 3. For Model 2, the version of split HMC without FlyMC performs better. For Model 3, the FlyMC version is the better performer.

\begin{figure}[!h] 
\centering
\input{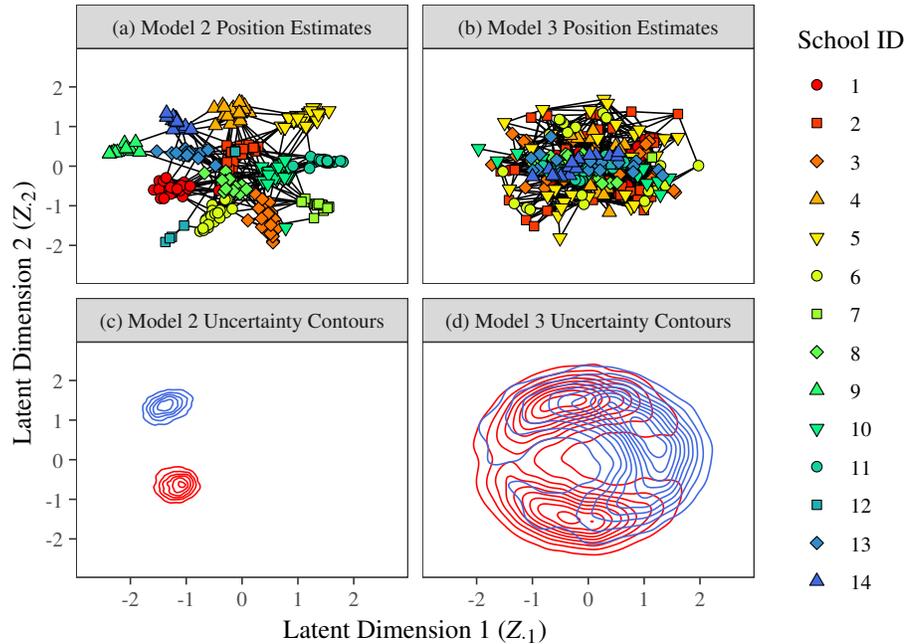}
\caption{The top panels depict point estimates of the teachers' latent positions for Models 2 and 3 (i.e. the one year, all schools models both with and without the same school covariate). These point estimates are obtained via multi-dimensional scaling on the posterior expectation of the matrix of squared latent distances. The shape/color combinations are assigned according to the teacher's school. The edges indicate information-sharing relationships. The bottom two panels depict posterior distribution contours for the latent positions for two teachers (one from school 1, one from school 14).} \label{experiment2Networks} 
\end{figure}

Both of these models are fit to the same network, but they differ in that Model 2 does not use the same school indicator in its covariate. As such, their disparity provides a good case study for when FlyMC is most useful. Table~\ref{design3} shows that for Model 3, $\hat{\tau}_{\text{EAP}} = 0.86$ for within-school dyads and $\hat{\tau}_{\text{EAP}} = 0.01$ for between-school dyads. This stark difference shows that the covariate in the link function has captured a sparsity of between-school edges. Meanwhile, Model 2 does not have access to the the same-school covariate, forcing it to use a single $\hat{\tau}_{\text{EAP}} = 0.43$ across all dyads. To compensate, Model 2's fitted link function is much steeper that of Model 3--- the $\hat{\gamma}_{\text{EAP}}^2$ values are 0.15 and 1.18, respectively. Turning our attention to Figure~\ref{experiment2Networks}, we can see how and why this compensation works.

Figures~\ref{experiment2Networks}(a) and (b) summarize the estimated latent positions in Models 2 and 3, respectively, with Figures~\ref{experiment2Networks}(c) and (d) illustrating the contours of the latent position posterior distributions for two individuals across the two models. Figure~\ref{experiment2Networks}(a) shows that Model 2's estimated latent positions are essentially acting as proxies for the absent same school covariate information, clustering the individuals in the latent space according to their school memberships. The steep decline of the link function thus allows it to capture the sparsity of between-school edges, with the few between-school ties in the network determining the relative positioning of the different school clusters. Meanwhile, the latent positions in Figure~\ref{experiment2Networks}(b) shows much less structure, with the posteriors in Figures~\ref{experiment2Networks}(d) being far more diffuse than those in \ref{experiment2Networks}(d). This suggests that the majority of the network's structure is already captured by the covariate in Model 3. Accordingly, the edge probabilities decay more gradually in the latent space.

The difference in algorithm performances between Models 2 and 3 thus boils down to the difference between $\tau$-driven sparsity versus $\gamma^2$-driven sparsity. The superior performance of the FlyMC algorithm for Model 3 is due to its ability exploit the sparsity (captured by $\tau$) between schools. The corresponding $\theta$ variables drastically downsample the number of dyads involved in each likelihood and gradient calculation, speeding up computation. This illustrates that FlyMC is especially useful when applied to networks containing mainy dyads for which the value of $\tau_{x_{ij}}$ is expected to be small. On the other hand, in models where the sparsity is driven $\gamma^2$, such as Model 2, there is less benefit to applying FlyMC.

We now shift our focus to Models 5 and 6, which have been fit to large longitudinal networks. For these models, the particularly strong performance of the split HMC algorithms compared to Metropolis within Gibbs can be attributed to two factors: the size of the network, and the extra structure imposed by the temporal autocorrelation in latent positions. The joint gradient-informed update of all nodes in split HMC is particularly well-suited to account for the autocorrelation in the prior, where the uninformed random walk update of Metropolis within Gibbs is not.

For Model 6, we see in Table~\ref{design2} that the same school covariate once again captures most of the network's sparsity by allowing for $\hat{\tau}_{\text{EAP}} = 0.01$ in across-school dyads that did not previously have an edge. However, FlyMC does not appear to be as beneficial in this case. This is due partially to the slower mixing of the $\tau$ variables in the FlyMC implementations---although they facilitate a closed-form Gibbs update, the auxiliary $\theta$ variables can also lead to extra autocorrelation in the chains for $\tau$ as they slowly evolve. Another contributor to the smaller disparity is a larger gap between tuned values of $\epsilon$. Our approach for tuning $\epsilon$ and $L$ yielded $\epsilon = 0.31$, $L = 7$ for the non-FlyMC implementation, and $\epsilon=  0.14$ and $L = 15$ for the FlyMC implementation. In the corresponding one year models, the gap was not as large--- $\epsilon = 0.22$, $L = 10$ for the non-FlyMC implementation compared to $\epsilon=  0.14$ and $L = 15$ for the FlyMC implementation.

\section{Concluding Remarks}

In this article we developed new algorithms for estimating Gaussian latent position models (GLPM, Equation \ref{gaussianlink}) using a variation of split Hamiltonian Monte Carlo \citep{shahbaba2014split} adapted to the link function of the model and a variation of Firefly Monte Carlo \citep{maclaurin2015firefly} for subsampling non-edge dyads while keeping the Monte Carlo algorithm exact up to Monte Carlo error. We conducted two empirical studies to investigate the performance of the split HMC and split HMC + FlyMC approaches: one on synthetic data (Section~\ref{exp1}) and one on real data concerning information-sharing among teachers and staff (Section~\ref{exp2}). 

Our synthetic data study demonstrated that when split HMC and split HMC + FlyMC are tuned to have an acceptance rate of around 0.8-0.85 (for integration time $T \approx 2$), these algorithms vastly outperform similarly well-tuned competitors in the literature, especially for large networks. The competitors we considered include standard algorithms such as Metropolis within Gibbs and elliptical slice sampling, as well as more sophisticated HMC algorithms such as the NUTS and Stan that adaptively set $T$. Across the board, our implementations of split HMC and split HMC + FlyMC outperformed all competitors, even the more sophisticated HMC algorithms. Of the two new algorithms proposed in this article, our implementation of split HMC seemed to be the better performer for denser networks, as well as smaller sparse networks for which the link function parameters were poorly identified. Split HMC + FlyMC performs best on large, sparse networks that contain sufficient information to identify the link function parameters. Given these results, we would expect split HMC + FlyMC to perform especially well in settings for which $\tau$, the maximum link probability in Equation~\ref{gaussianlink}, is very small, such as the sparse graphon version of GLPMs \citep{borgs2014sparse, spencer2017projective}.

At first, we were surprised by the extent to which the HMC integration time $T$ at approximately 2 outperformed the adaptive strategies for setting $T$ used in NUTS and Stan. Adaptive strategies have been shown to outperform fixed integration strategies across a variety of models and perform at least comparably for others \citep{hoffman2014no, betancourt2016identifying}. However, the relatively poor performance of these algorithms on LPMs is less surprising when one considers the criteria used by Stan and NUTS to choose $T$. Both algorithms are configured to optimize the mixing of the latent positions. However, as we elaborate on in Section~\ref{ESS} of the Appendix, the latent positions themselves are under-identified in the posterior---the edge probabilities or latent distances are more appropriate targets because they are identified in the model. This insight suggests that adapting the NUTS criterion to optimize the mixing of functions of the parameters could be a worthwhile extension.

Our real data study demonstrated that FlyMC also performs well when $\tau$ varies with a categorical covariate, taking on small values for some categories. Another potential avenue of future research would be to consider a hybrid of split HMC and split HMC + FlyMC. FlyMC would only be applied to the subset of dyads for which $\tau$ is expected to be small. FlyMC would thus be exploited only where it is most effective, limiting any potential mixing problems due to the extra auxiliary variables.

In our real data study, we also considered the application of our algorithm to fit Gaussian LPMs for longitudinal networks. Of the variety of different ways to configure longitudinal LPMs \citep{kim2018review}, we used a simple model structure that used a structured Gaussian prior to promote serial dependence of the nodes' latent positions across time points, as well as covariates to promote persistent edges across time points. For these models, our split HMC algorithm performed particularly well because gradient-informed proposals naturally accommodate the extra structure in the prior. Therefore, we anticipate that our insights could also be applied to achieve substantial computational gains in other more structured priors for LPMs, such as latent cluster model \citep{krivitsky2009representing} or the multi-resolution network model \cite{fosdick2018multiresolution}. Recently, \cite{turnbull2020advancements} explored the use of sequential Monte Carlo \citep{doucet2009tutorial} for Bayesian inference of longitudinal latent position models. They found that the algorithm scaled poorly to large networks due to expensive likelihood evaluations. These findings suggest that a combination of their approach with the algorithms we present here could be fruitful.

Finally, it is worth briefly commenting on our motivation for focusing on the Gaussian link function instead of the traditional logistic link function. When proposing the GLPM, \cite{rastelli2016properties} argued that the Gaussian LPM yields results that are analogous and comparable to those of the logistic LPM, whilst providing interpretable link function parameters and making it easier to derive theoretical results. Since then, \cite{spencer2017projective} proved consistent estimation results for the Gaussian link function---analogous results under the logistic link function remain an open problem. In this article, we have demonstrated additional benefits of the Gaussian link function over the logistic link function. The closed-form Gaussian decomposition underlying our split HMC algorithm is not possible with a logistic link, and the FlyMC strategy cannot be applied because it relies on a factorization of the sparsity parameter $\tau$ from the link function. Moreover, the likelihood of the logistic link LPM is not differentiable when two nodes have the same position, which complicates the application of HMC. For all of these reasons, we propose that the Gaussian LPM would be a more suitable ``default'' choice of the link function---the logistic link should only be used when its particular functional form is justified by the application.

\section*{Acknowledgements}
The authors have no conflicts of interest to disclose.
%
%
%

\bibliographystyle{apa-good}

\bibliography{Bibliography-MM-MC}

\section{Appendix}

\subsection{Computational Details of Experiments} \label{compdetails}

The implementations of all of the algorithms we used for the empirical studies are available as part of an R package. The Metropolis within Gibbs algorithms use a multivariate uniform distribution over $[z_i - \delta, z_i + \delta]$ as the proposal distribution $q_{\delta}$, with $\delta$ tuned to target an acceptance rate within 20 and 30 percent. Similarly, the value of $\epsilon$ for the split HMC algorithms was tuned to target an acceptance rate between 80 and 85 percent. The tuned values of $\epsilon$ and $\delta$ for all of the configurations in Study 1 are available in Figure~\ref{adfigs}, as well as Table 1 of the main text.

All experiments (except those involving Stan) were run using the Bridges High Performance Computing System \citep{nystrom2015bridges} at the Pittsburgh Supercomputing Center. The computing costs were supported by XSEDE Integrated Advanced Digital Services \citep{towns2014xsede}. Because of a software version incompatibility issue, we had to instead run the Stan experiments in a different cluster computing environment called Hydrea. Timing tests revealed that the Bridges supercomputer runs roughly 3.3 to 3.6 times slower than analogous runs on the Hydra computing cluster. To facilitate direct comparisons between the cluster and Bridges experiments, all run times of the Stan algorithms were multiplied by a factor of 3.3 when determining the comparisons shown in Figure~\ref{experiment1afig}. Version 2.18.2 of \texttt{rstan} was used to run the experiment, and \texttt{coda} version 0.19-2 was used when calculating effective sample sizes.

\subsection{Details of Tuning the Algorithms}\label{tuning}

\begin{figure}[!h] 
\centering
\resizebox{\textwidth}{6.2cm}{
\input{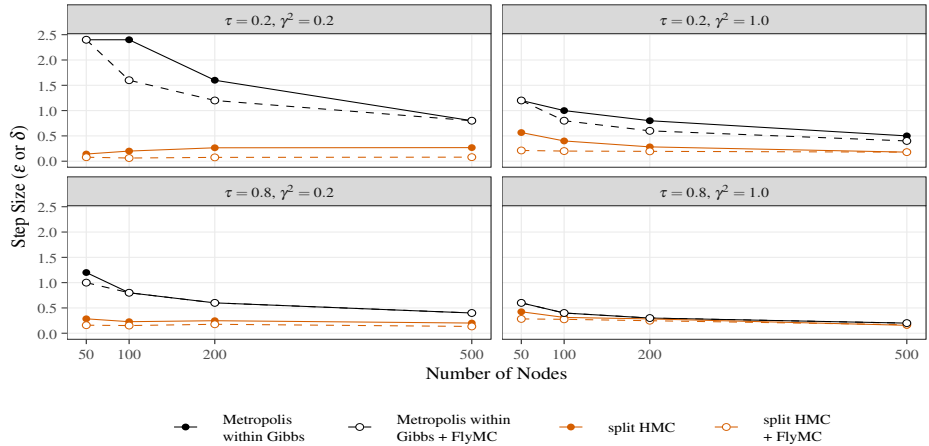}
}
\caption{Four panels depicting the tuned step size parameters used by Metropolis within Gibbs, split HMC, and their FlyMC counterparts for fitting the 16 different networks considered in Study 1. Each panel displays the step size parameter ($\delta$ for Metropolis methods and $\epsilon$ for HMC methods) used for the 50, 100, 200, and 500 node networks generated using the parameter values featured in the panel heading. Point color, point shape, line color, and line shape are used to distinguish the algorithms.} \label{tuningeps} 
\end{figure}

Where appropriate, we tuned the parameters of the algorithms to promote efficient computation. Stan has a sophisticated (and computationally intensive) tuning strategy for choosing its step size $\epsilon$ along with a mass matrix $M$. For details, see \cite{carpenter2017stan}. For all of the non-Stan algorithms, we used a light tuning strategy based on a sequence of short (100 iteration) preliminary runs to iteratively select reasonable values for the parameters. For the updates of $Z$, the Metropolis and Metropolis + FlyMC step sizes were chosen to target an acceptance rate in the range $[0.2,0.3]$. For the split HMC and split HMC + FlyMC algorithms, the value of $\epsilon$ was determined by targeting an acceptance rate within $[0.8, 0.85]$. The value of $L$ was chosen simultaneously to ensure $T = L \epsilon \approx 2$.  We have found through a wide array of preliminary experiments that these values tend to give results that are close to optimal without taking too much tuning time. For the NUTS algorithms, the value of $\epsilon$ chosen for the analogous split HMC algorithm was used. The elliptical slice algorithms have no tuning parameters. The step-size for the Metropolis $\tau$ updates was also chosen based on short preliminary runs, targeting an acceptance rate in the range $[0.2,0.3]$. The step sizes used to update $Z$ for split HMC, split HMC + FlyMC, Metropolis within Gibbs, and Metropolis within Gibbs + FlyMC are provided in Figure~\ref{tuningeps} in Section~\ref{adfigs}.

\subsection{Measuring relative efficiency of MCMC Algorithm for LPMs} \label{ESS}
There are two principal criteria by which to judge the efficiency of a Markov chain for approximating a posterior distribution.

\begin{enumerate}
\item How efficiently does the MCMC sequence approximate the posterior expectation of a desired function of the parameters?
\item What is the computational expense of generating this chain?
\end{enumerate}
Simple metrics exist for gauging each of these criteria, which we describe below.

For criterion 1, it is well-known that that mean estimates stemming from a geometrically ergodic \citep{roberts1997geometric} Markov chain are subject to a Markov chain central limit theorem (CLT) \citep{tierney1994markov}. Analogous to the standard (independent samples) central limit theorem, for which the variance of the estimator is inversely proportional to the raw sample size, the variance in the Markov chain CLT is inversely proportional to the \emph{Effective sample size} \citep{kass1998markov, ripley2009stochastic}.

Letting $\theta^1, \theta^2, \ldots, \theta^N$ denote $N$ draws from a Markov chain and $f$ denote an arbitrary function, the effective sample size $ESS_f$ for estimating the expectation of $f(\theta)$ is defined as 
\[
\textrm{ESS}_{f}(\theta^1,\ldots, \theta^N) = \frac{N}{1 + 2 \sum_{t=1}^{\infty} \rho_{t,f}}
\]
where 
\[
\rho_{t,f} = \frac{\int f(\theta^{i+t}) f(\theta^{i}) p(\theta^{i}) \textrm{d} \theta^{i}}{\int f(\theta^{i})^2 p(\theta^{i}) \textrm{d} \theta^{i}}
\]
denotes the $t$-lag autocorrelation of the function $f(\theta)$ in the Markov chain. 

The effective sample size thus takes into account all possible lags of autocorrelations, recognizing that highly autocorrelated chains represent fewer bits of independent information than an uncorrelated analog. In practice, estimating the effective sample size of a chain is done by estimating the autocorrelations. For our purposes, we use the effective sample size estimator implemented in the R package \texttt{coda} \citep{plummer2006coda}. 

We should note that we have not formally proved that our proposed HMC + FlyMC algorithm produces a geometrically ergodic Markov chain (see \cite{mangoubi2017rapid, livingstone2019geometric, mangoubi2019mixing} for recent progress on related problems). However, the effective sample size remains an intuitive metric for the efficiency of a Markov chain approximation, due to its penalization of autocorrelations. Moreover, conservative confidence intervals based on the effective sample size can still be constructed even when geometric ergodicity does not necessarily hold \citep{rosenthal2017simple}. For this reason, we feel it is still a suitable metric to use when comparing chains. 

To assess criterion 2 (the computational expense of generating a chain), we simply measure the runtime\footnote{None of the algorithms we consider here are memory-intensive. If any were, it might also make sense to report the memory requirements.} of the algorithm. Although there is opportunity for parallelization in some of the algorithms (e.g. simultaneous updating of FlyMC variables), we do not take advantage of such opportunities---all our implementations perform the various steps in series. 

Overall, both criterion 1 and criterion 2 should be considered simultaneously when judging the efficiency of a Markov chain---a fast MCMC algorithm is not necessarily accurate, and an accurate MCMC algorithm is not necessarily fast. The most popular metric for combining these criteria is the effective sample size per second \citep{gamerman2006markov}, defined by
\[
\textrm{ESS per second} = \frac{\textrm{ESS}_{f}(\theta^1,\ldots, \theta^N)}{\textrm{time (in seconds) taken to compute the chain }  \theta^1,\ldots, \theta^N},
\]
sometimes referred to as Markov chain efficiency \citep{web}. This metric provides a straightforward way to compare the performance of two MCMC algorithms. If one MCMC algorithm produces twice as many effective samples per second as another, that means it is twice as efficient, obtaining an equally accurate approximation of a desired posterior expectation in roughly half the time. This begs the question---what posterior expectation do we desire? 

Typically, the posterior mean of the various parameters is the natural choice \citep{carpenter2017stan}. However, for the LPM, the posterior mean of each latent positions is inappropriate. Due to the invariance of the likelihood under isometric transformations (e.g. rotations and reflections \citep{shortreed2006positional}), all latent positions are guaranteed to have posterior mean of 0. If our target were the posterior means of the positions, MCMC would be unnecessary.

A reasonable target should instead be well-identified in the model, such as the probabilities of edges between the nodes. These values depend both on the latent positions and the parameters of the link function. However, there are $n(n-1)/2$ such probabilities in total. For large networks, computing the effective sample size for all of them is computationally burdensome. Moreover, many of these expected probabilities will be very close to 0 in sparse networks, creating numerical underflow problems in practice.

To avoid having to compute the Markov efficiency for all $n(n-1)/2$ unique pairs in $[n]^2$, we instead consider a uniformly random subset of 500 dyads (sampled without replacement) for each of our empirical studies. Subsampling drastically reduces the amount of computation required while still providing a summary of how well the chain is mixing for all nodes. To avoid the underflow problems associated with estimating the raw probabilities, we calculate the effective sample size of estimating the log probabilities instead. That is, for a given $i,j \in [n]^2$, we consider 
\[
f(Z, \gamma, \tau,X) = \log(\tau_{x_{ij}}) - \frac{\|z_i - z_j\|^2}{2}.
\]
as the function for which we calculate the posterior expectation. These functions are much more numerically tractable than the raw probabilities, whilst preserving strong ties with other quantities of interest (i.e. the distance between nodes and the density parameter $\tau$). 

Finally, we can now bring all of this together to define an interpretable quantity for reporting the relative efficiency of MCMC algorithms. Metropolis within Gibbs is currently the most popular MCMC algorithm for posterior computation of LPMs, making it the natural choice for the baseline against which to compare other algorithms in our empirical studies. Accordingly, we report each algorithm's efficiency relative to Metropolis within Gibbs for each subsampled dyad. That is, for a chain $\theta^1, \ldots, \theta^N$, we calculate
\begin{align*}
\frac{\textrm{ESS}_{f}(\theta^1,\ldots, \theta^N)}{\textrm{ESS}_{f}(\theta_m^1,\ldots, \theta_m^N)} \times \frac{\textrm{time (in seconds) taken to compute the chain }  \theta_m^1,\ldots, \theta_m^N}{\textrm{time (in seconds) taken to compute the chain }  \theta^1,\ldots, \theta^N} \label{comparison}
\end{align*}
for each dyad, where $\theta_m^1, \ldots, \theta^N_m$ denotes draws according to a well-tuned Metropolis within Gibbs algorithm exploring the same posterior.

\subsection{Full Conditional Distributions} \label{math}

Sections~\ref{FlyMCmath} and \ref{splitmath} provide the details of the targeted joint distribution as well as the relevant conditional distributions and algorithmic steps for split HMC + FlyMC and split HMC, respectively.

\subsubsection{Split HMC + FlyMC}\label{FlyMCmath}

The full joint distribution of all observed data and parameters for our split HMC + FlyMC strategy can be decomposed as
\begin{align}
p(A, U, Z, \theta, \tau, \gamma^2 \mid x, a_*, b_*, \alpha, \beta, \Omega) &= p(U| \gamma^2, A, \Omega) p(A \mid Z, \theta, \tau, \gamma^2, x) p(Z\mid \gamma^2, \Omega)\\
&\times p(\theta \mid \tau, x) p(\tau \mid \alpha, \beta) p(\gamma^2 \mid a_*,b_*),
\end{align}
where $x \in [C]^{n \times n}$ denotes the observed covariates, $a_*,b_* \in \mathbb{R}_+$ denote the hyperparameters for the inverse gamma prior on $\gamma^2$, $\alpha, \beta \in \mathbb{R}_+^C$ denote the hyperparameters for the beta prior(s) on $\tau$, and $\Omega$ denotes the prior covariance for the latent positions $Z$ before the re-parametrization. The full expression for each of the components in the decomposition is
\begin{align*}
p(A \mid Z, \theta, \tau, x) &= \left(\prod_{(i,j) \in E_A} \tau_{x_{ij}} \right) \exp{\left(-\frac{1}{2}\sum_{\ell = 1}^d Z_{\cdot \ell}^T L_A Z_{\cdot \ell} \right)}\prod_{\substack{\theta_{ij} = 1 \\ A_{ij} = 0}} \left(1 - \exp{\left(-\frac{1}{2} \|z_i - z_j\|^2 \right)} \right)\\
p(Z\mid \gamma^2, \Omega) &= \frac{1}{(2 \pi)^{nd/2} \gamma^{nd} \text{det}\left(\Omega\right)^d}\exp{\left(- \frac{1}{2 \gamma^2} \sum_{\ell = 1}^d Z_{\cdot \ell}^T \Omega^{-1} Z_{\cdot \ell} \right)}\\
p(\gamma^2 \mid a_*,b_*) &= \frac{b_*^{a_*}}{\Gamma(a_{*})} (\gamma^{2})^{-a_{*}-1} \exp{\left(-\frac{b_{*}}{\gamma^2} \right)}\\
p(\theta \mid \tau, x) &= \prod_{(i,j) \in A} \tau_{x_{ij}}^{\theta_{ij}} \left(1 - \tau_{x_{ij}}\right)^{1 - \theta_{ij}}\\
p(\tau \mid \alpha, \beta) &= \prod_{c = 1}^C \frac{\Gamma(\alpha_c + \beta_c)}{\Gamma(\alpha_c) \Gamma(\beta_c)}\tau_{c}^{\alpha_c - 1} \left(1- \tau_{c}\right)^{\beta_c - 1}\\
p(U| \gamma^2, A, \Omega) &= \frac{\text{det}\left(\frac{1}{\gamma^2}\Omega^{-1} + L_A\right)^d}{(2 \pi)^{nd/2}} \exp{\left(- \frac{1}{2} \sum_{\ell = 1}^d U_{\cdot \ell}^T \left(\frac{1}{\gamma^2}\Omega^{-1} + L_A\right)^{-1} U_{\cdot \ell} \right)}
\end{align*}
where $\text{det}(\cdot)$ denotes the determinant of a matrix and $\Gamma(\cdot)$ denotes the Gamma function. To perform MCMC on this distribution, we alternate through the following conditional updates
\begin{enumerate}
\item Use split HMC described in Algorithm 1 to update $(Z,U)$ according to the conditional posterior density $p(U,Z \mid A, \theta, \gamma^2, \Omega)$ defined by
\begin{align*}
p(U,Z \mid A, \theta, \gamma^2, \Omega) &\propto \exp{\left(- \frac{1}{2} \sum_{\ell = 1}^d Z_{\cdot \ell}^T \left(\frac{1}{\gamma^2}\Omega^{-1} + L_A\right) Z_{\cdot \ell} + U_{\cdot \ell}^T \left(\frac{1}{\gamma^2}\Omega^{-1} + L_A\right)^{-1} U_{\cdot \ell} \right)}\\
& \times \prod_{\substack{\theta_{ij} = 1 \\ A_{ij} = 0}} \left(1 - \exp{\left(-\frac{1}{2} \|z_i - z_j\|^2 \right)} \right)
\end{align*}
Note that in this case, the mass matrix for HMC is given by 
\begin{align*}
M &= \left(\frac{1}{\gamma^2}\Omega^{-1} + L_A\right)
\end{align*}
which amounts to having it adaptively updated according to $\gamma^2$.
\item Apply the Metropolis-Hastings to update each of $\theta_{ij}$ for which $A_{ij} = 0$ according to the conditional posterior density $p(\theta \mid \tau, x)$ defined by 
\begin{align*}
p(\theta_{ij} = 0 \mid A_{ij} =0, \tau_{x_{ij}}) &= \frac{1 - \tau_{x_{ij}}}{1 - \tau_{x_{ij}} \exp{\left( - \frac{1}{2}\|z_i - z_j \|^2 \right)}}.
\end{align*}
Recall that because $p(\theta_{ij} = 0 \mid A_{ij} =1, \tau_{x_{ij}}) = 0$, the $\theta_{ij}$ for which $A_{ij} = 1$ need not be updated---they are known to be fixed at one.
\item Apply a Gibbs updates to update each entry in $\tau$ according to the beta conditional posterior densities
\begin{align*}
p(\tau_c | \theta, \alpha_c, \beta_c) &= \frac{\Gamma(\alpha_c + \beta_c + \Theta_c^0 + \Theta_c^1)}{\Gamma(\alpha_c + \Theta_c^0) \Gamma(\beta_c + \Theta_c^1)} \tau_c^{\alpha_c + \Theta_c^1 - 1} \left(1 - \tau_c\right)^{\beta_c + \Theta_c^0 -1}
\end{align*}
where $\Theta_0^c$ and $\Theta_1^c$ are defined as in the main text, reproduced below for easy access.
\begin{align*}
\Theta^0_c = |\left\{ (i,j) \in [n]^2: \theta_{ij} = 1 \text{ and } x_{ij} = c \right\}|\\
\Theta^1_c = |\left\{ (i,j) \in [n]^2: \theta_{ij} = 0 \text{ and } x_{ij} = c \right\}|.
\end{align*}
\item Apply a Gibbs update to $\gamma^2$ according to the inverse gamma conditional density
\begin{align*}
p(\gamma^2 |a_*,b_*, Z, \Omega ) &=  \frac{\left(b_* + \frac{1}{2} \sum_{\ell = 1}^d Z_{\cdot \ell}^T \Omega^{-1} Z_{\cdot \ell} \right)^{a_* + \frac{nd}{2}}}{\Gamma(a_* +  \frac{nd}{2}) \left(\gamma^{2}\right)^{a_*+\frac{nd}{2}+1}}  \exp{\left(-\frac{\left(b_* + \frac{1}{2} \sum_{\ell = 1}^d Z_{\cdot \ell}^T \Omega^{-1} Z_{\cdot \ell} \right)}{\gamma^2} \right)}.
\end{align*}
Note that this expression above arises only after marginalizing the momentum variables $U$. Typically, after such a marginal update in MCMC, the $U$ parameter would need to be updated according to its conditional distribution. In practice, this is not necessary, as $U$ is not one of the target parameters (moreover, the Gibbs update is immediately applied again in the following Step 1).
\end{enumerate}

\subsubsection{Split HMC}\label{splitmath}

The full joint distribution of all observed data and parameters for our split HMC strategy can be decomposed as
\begin{align*}
p(A, U, Z, \tau, \gamma^2 \mid x, a_*, b_*, \alpha, \beta, \Omega) &= p(U| \gamma^2, A, \Omega) p(A \mid Z, \tau, \gamma^2, x) p(Z\mid \gamma^2, \Omega)\\
& \times p(\tau \mid \alpha, \beta) p(\gamma^2 \mid a,b),
\end{align*}
where $x \in [C]^{n \times n}$ denotes the observed covariates, $a_*,b_* \in \mathbb{R}_+$ denote the hyperparameters for the inverse gamma prior on $\gamma^2$, $\alpha, \beta \in \mathbb{R}_+^C$ denote the hyperparameters for the beta prior(s) on $\tau$, and $\Omega$ denotes the prior covariance for the latent positions $Z$ before the re-parametrization. The full expression for each of the components in the decomposition is
\begin{align*}
p(A \mid Z, \tau, \gamma^2, x) &= \left(\prod_{(i,j) \in E_A} \tau_{x_{ij}} \right) \exp{\left(-\frac{1}{2}\sum_{\ell = 1}^d Z_{\cdot \ell}^T L_A Z_{\cdot \ell} \right)}\prod_{(i,j) \notin E_A} \left(1 - \tau_{x_{ij}}\exp{\left(-\frac{1}{2} \|z_i - z_j\|^2 \right)} \right)\\
p(Z\mid \gamma^2, \Omega) &= \frac{1}{(2 \pi)^{nd/2} \gamma^{nd} \text{det}\left(\Omega\right)^d}\exp{\left(- \frac{1}{2 \gamma^2} \sum_{\ell = 1}^d Z_{\cdot \ell}^T \Omega^{-1} Z_{\cdot \ell} \right)}\\
p(\gamma^2 \mid a_*,b_*) &= \frac{b_*^{a_*}}{\Gamma(a_*)} (\gamma^{2})^{-a_*-1} \exp{\left(-\frac{b_*}{\gamma^2} \right)}\\
p(\tau \mid \alpha, \beta) &= \prod_{c = 1}^C \frac{\Gamma(\alpha_c + \beta_c)}{\Gamma(\alpha_c) \Gamma(\beta_c)}\tau_{c}^{\alpha_c - 1} \left(1- \tau_{c}\right)^{\beta_c - 1}\\
p(U| \gamma^2, A, \Omega) &= \frac{\text{det}\left(\frac{1}{\gamma^2}\Omega^{-1} + L_A\right)^d}{(2 \pi)^{nd/2}} \exp{\left(- \frac{1}{2} \sum_{\ell = 1}^d U_{\cdot \ell}^T \left(\frac{1}{\gamma^2}\Omega^{-1} + L_A\right)^{-1} U_{\cdot \ell} \right)}
\end{align*}
where $\text{det}(\cdot)$ denotes the determinant of a matrix and $\Gamma(\cdot)$ denotes the Gamma function. To perform MCMC on this distribution, we alternate through the following conditional updates
\begin{enumerate}
\item Use split HMC described in Algorithm 1 to update $(Z,U)$ according to the conditional posterior density $p(U,Z \mid A, \gamma^2, \Omega)$ defined by
\begin{align*}
p(U,Z \mid A, \gamma^2, \Omega) &\propto \exp{\left(- \frac{1}{2} \sum_{\ell = 1}^d Z_{\cdot \ell}^T \left(\frac{1}{\gamma^2}\Omega^{-1} + L_A\right) Z_{\cdot \ell} + U_{\cdot \ell}^T \left(\frac{1}{\gamma^2}\Omega^{-1} + L_A\right)^{-1} U_{\cdot \ell} \right)}\\
& \times \prod_{(i,j) \notin E_A} \left(1 - \tau_{x_{ij}}\exp{\left(-\frac{1}{2} \|z_i - z_j\|^2 \right)} \right)
\end{align*}
Note that in this case, the mass matrix for HMC is given by 
\begin{align*}
M &= \left(\frac{1}{\gamma^2}\Omega^{-1} + L_A\right)
\end{align*}
which amounts to having it adaptively updated according to $\gamma^2$.
\item Apply a random walk Metropolis to update each entry in $\tau$ using its posterior conditional distribution
\begin{align*}
p(\tau_c \mid A, x_{ij}, \alpha_c, \beta_c) &\propto \tau_c^{\alpha_c + \zeta_c^1 - 1} (1- \tau_c)^{\beta_c - 1}  \prod_{\substack{x_{ij} = c \\ A_{ij} = 0}} \left(1 - \tau_{x_{ij}}\exp{\left(-\frac{1}{2} \|z_i - z_j\|^2 \right)} \right)
\end{align*}
where $\zeta^1_c$ is defined as
\begin{align*}
\zeta^1_c = |\left\{ (i,j) \in [n]^2: A_{ij} = 1 \text{ and } x_{ij} = c \right\}|.
\end{align*}
We recommend updating each entry individually, using a uniform proposal centered at its current value with step-size tuned to obtain an acceptance rate within 20 and 30 percent. This is the strategy we used throughout the article.
\item Apply a Gibbs update to update $\gamma^2$ according to the inverse gamma conditional density
\begin{align*}
p(\gamma^2 |a_*,b_*, Z, \Omega ) &=  \frac{\left(b_* + \frac{1}{2} \sum_{\ell = 1}^d Z_{\cdot \ell}^T \Omega^{-1} Z_{\cdot \ell} \right)^{a_* + \frac{nd}{2}}}{\Gamma(a_* +  \frac{nd}{2}) \left(\gamma^{2}\right)^{a_*+\frac{nd}{2}+1}}  \exp{\left(-\frac{\left(b_* + \frac{1}{2} \sum_{\ell = 1}^d Z_{\cdot \ell}^T \Omega^{-1} Z_{\cdot \ell} \right)}{\gamma^2} \right)}.
\end{align*}
Note that this expression above arises only after marginalizing the momentum variables $U$. Typically, after such a marginal update in MCMC, the $U$ parameter would need to be updated according to its conditional distribution. In practice, this is not necessary, as $U$ is not one of the target parameters (moreover, the Gibbs update is immediately applied again in the following Step 1).
\end{enumerate}

\subsection{Algorithm Performance as Latent Space Dimension Grows}

\begin{figure}[!h] 
\centering
\resizebox{\textwidth}{7cm}{
\input{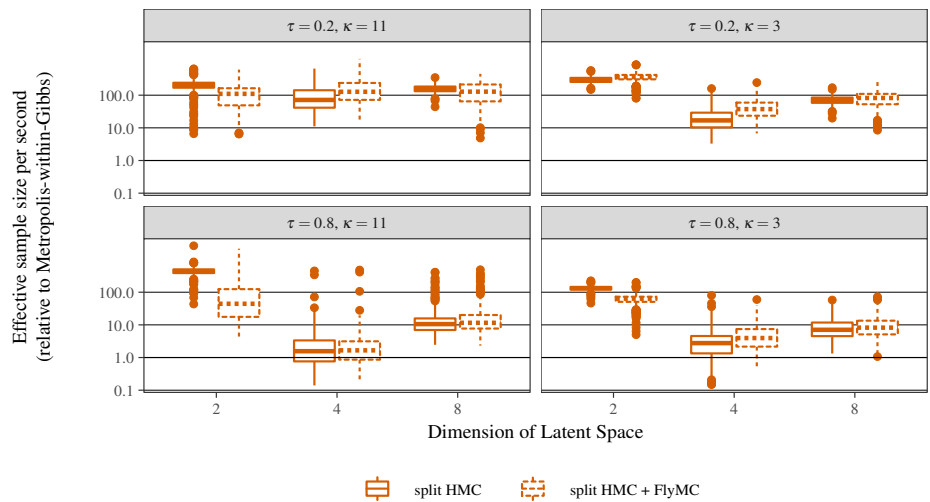} 
}
\caption{Boxplots showing the relative efficiency of Split HMC + FlyMC and Split HMC relative to Metropolis within Gibbs across 500 dyads in each network.} \label{experimentd}
\end{figure}

 Figure~\ref{experimentd} below is the analog to Figure~\ref{experiment1cfig} in the main text, showing how our algorithms perform for various dimensions of the latent space ($d = 2$, $d =4$, and $d= 8$). These results were obtained as a follow-up companion to the original Study 1, as suggested by an anonymous reviewer. All networks considered consist of $n=500$ nodes, with the simulation set-up otherwise similar to Study 1.
 
The main difference is that we did not hold $\gamma^2$ fixed as $d$ increased like we did for $n$ in Study 1. Recall that in Gaussian latent position model with unit isotropic Gaussian latent positions, the expected number of neighbours of a random node is given by
 \begin{align*}
\mathbb{E}(\sum_{i=1}^n A_{ij}) &= n \tau \left(1 + \frac{2}{\gamma^2} \right)^{-d/2}.
\end{align*}
As such, the expected density of the networks would change drastically as $d$. Instead, we adjusted $\gamma^2$ along with $d$ to ensure the quantity
\begin{align*}
\kappa = \left(1 + \frac{2}{\gamma^2} \right)^{d/2}
\end{align*}
stayed fixed for each network sparsity regime. We chose $\kappa = 3$ and $\kappa = 11$ because for $d=2$, these correspond to the values of $\gamma^2 = 1.0$ and $\gamma^2 = 0.2$ considered in Study 1. 

Inspecting Figure~\ref{experimentd}, there does not appear to be a clear upward trend in any of the regimes. Though the HMC-based methods still appear to outperform Metropolis within Gibbs, the difference does not appear to grow reliably with $d$. At first, this may seem to be a surprising result; HMC-based methods are known to perform especially well on high-dimensional posterior distributions, relative to random walk-based methods. For this reason, one might expect our split HMC algorithms to perform relatively better as $d$ increased, much like they did as $n$ increased.

However, there is a difference in geometry between the growing $n$ regime and the growing $d$ regime for LPMs. Though its size stays fixed with $n$, the class of isometric transformations under which the posterior distribution remains invariant grows with $d$. As such, posteriors over higher dimensional latent positions tend to be less constrained, with the latent positions themselves being poorly identified. It is plausible that the benefits of the gradient information is dampened as $d$ grows. At the very least, the benefits do not appear to increase with $d$ in this study.



\subsection{Additional Figures and Tables}\label{adfigs}

\begin{figure}[!h] 
\centering
\resizebox{\textwidth}{6.2cm}{
\includegraphics{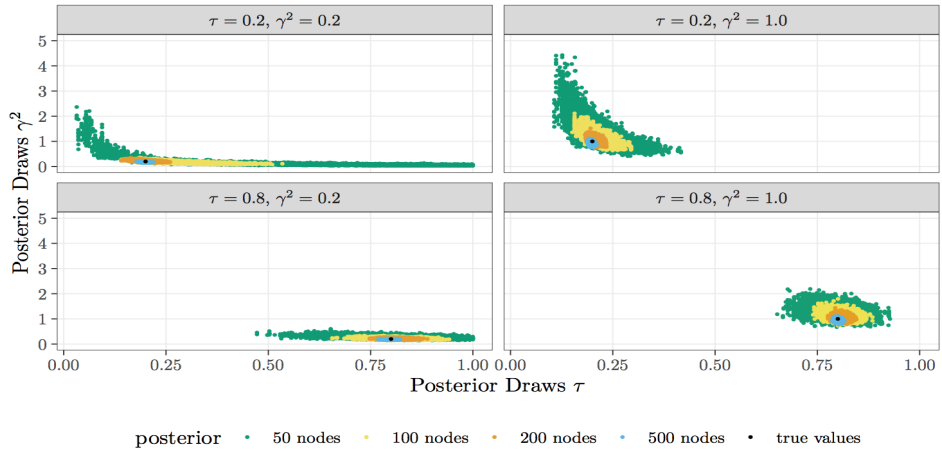}
}
\caption{Four panels depicting the marginal joint posterior of $\tau$ and $\gamma^2$ for the 16 different networks considered in Study 1. Each panel displays draws from the joint posterior for the 50, 100, 200, and 500 node networks generated using the panel heading parameter values. The draws for different sized networks are differentiated by color, with a black point used to indicate the ground truth values of $\tau$ and $\gamma^2$.} \label{tauvsgamma} %
\end{figure}
\end{document}